\documentclass[twocolumn,trackchanges,twocolappendix]{aastex631}

\accepted{for publication in the Astrophysical Journal on 24 May 2025 }

\begin{document}

\title{Changing Look AGN: A study of Optical/UV and the Highly Ionized Fe K$\alpha$ X-ray Line Flux Variations Using Photo-Ionization Simulations}

\author[0000-0003-4586-0744]{Tek P. Adhikari}
\affiliation{CAS Key Laboratory for Research in Galaxies and Cosmology, Department of Astronomy, University of Science and Technology of China, Hefei, Anhui 230026, China}
\affiliation{School of Astronomy and Space Science, University of Science and Technology of China, Hefei, Anhui 230026, China}
\email{tek@ustc.edu.cn}

\author[0000-0003-0793-6066]{Santanu Mondal}
\affiliation{Indian Institute of Astrophysics, 2nd Block Koramangala, Bangalore 560034, Karnataka, India}

\author[0000-0003-3667-1060]{Zhicheng He}
\affiliation{CAS Key Laboratory for Research in Galaxies and Cosmology, Department of Astronomy, University of Science and Technology of China, Hefei, Anhui 230026, China}
\affiliation{School of Astronomy and Space Science, University of Science and Technology of China, Hefei, Anhui 230026, China}
\email{zcho@ustc.edu.cn}

\author{Agata Rozanska}
\affiliation{Nicolaus Copernicus Astronomical Center of the Polish Academy of Sciences (CAMK PAN), Bartycka 18, 00-716, Warsaw, Poland}
\author{Krzysztof Hryniewicz}
\affiliation{National Centre for Nuclear Research, Astrophysics Division, Pasteura 7, 02-093 Warsaw, Poland}

\begin{abstract}
Significant variability in broad emission line strengths of active galactic nuclei (AGN) over months to years has been observed, often accompanied by intrinsic continuum changes. Such spectral variability challenges the traditional AGN classification scheme, which attributes differences between Type 1 and Type 2 to geometrical effects, as transitions between these types occur on timescales shorter than viscous ones. In this work, using the {\sc cloudy} photo-ionization simulations, we investigated the response of the major emission line fluxes, in the optical/UV and hard X-ray bands, to changes in the intensity and shape of the continuum emission of the AGN under two scenarios: (i) changes in the X-ray power-law while keeping disc emission fixed, and (ii) broadband continuum variations. We demonstrate that BLR line fluxes are insensitive to X-ray power-law changes alone. Considering a well-studied case of the changing-look (CL) AGN Mrk 1018, which exhibits variations in the intrinsic disc emission, as well as the X-ray power-law, our simulations reproduce observed brightening and dimming trends of the BLR emission. Moreover, we show that the highly ionized Fe K$\alpha$ X-ray flux, primarily produced by the H-like and He-like ions of Fe, strongly depends on the X-ray strength of the intrinsic SED. These findings suggest that the origin of highly ionized Fe K$\alpha$ emission is in the coronal part of the accretion disk and that the CL phenomenon can be triggered by intrinsic changes in the accretion properties of AGN.
\end{abstract}


\section{Introduction}
Spectral variation in active galactic nuclei (AGN) across the wavelength bands is a universal phenomenon. The continuum variation in optical/UV and X-ray wavelengths for AGN has been studied in detail in the literature \citep{Nandra1998,Zetzl2018,Hagen2024}. The correlation between continuum emission at different wavelength bands (optical/UV and X-rays) has been extensively studied in the past, thanks to the availability of multi-wavelength data from state-of-the-art instruments. These studies often show that optical/UV and X-ray emissions are not as well correlated \citep{Hagen2024} as expected from models of X-ray production and the origins of UV emission.

The basic variability timescales of AGN spectral emission, predicted by the standard theory of accretion discs around black holes—i.e., the Shakura-Sunyaev disk \citep{Shakura1973}—should be comparable to viscous timescales, on the order of several thousand years. However, much more complex spectral variability has been observed in the optical/UV emission lines over timescales of just a few months to years (e.g., \citet{Macleod2010}). This rapid variability, far shorter than the viscous timescale, is often attributed to intrinsic changes in the accretion disk itself \cite[e.g.,][]{Noda2018,Ricci2020,Sniegowska2020A&A,Mondal2022,Ricci2023,Veronese2024,Jana2024}.

Optical and UV emission lines are critical in characterizing the properties and composition of the AGN environment. These lines are primarily attributed to the photo-ionization process, where the continuum radiation emitted by the accretion disk photo-ionizes gas clouds located at sub-parsec to parsec scales, producing the observed emission features. The emission lines play a key role in classifying AGN types, a distinction that largely depends on the observer’s viewing angle \citep{Antonucci1993,Urry1995,Beckmann2012}. AGNs are traditionally categorized as Type 1 or Type 2 based on the presence or absence of broad emission line components. However, recent observations reveal that some AGNs can transit between these types over short timescales, a phenomenon referred to as "Changing-Look" (CL) behavior. This variability, characterized by significant changes in the strength of broad emission lines, challenges the traditional geometry-dependent classification model.

A fundamental question arises: can the CL behavior of AGNs be linked to their X-ray emission properties? If CL behavior originates from changes in the intrinsic properties of the accretion flow, it is reasonable to expect corresponding variations in the X-ray continuum emission. Recent studies have indeed observed correlations between the CL phenomenon and X-ray continuum variability \citep{LaMassa2017,Oknyansky2019MNRAS,Kollatschny2020,Guolo2021}. For instance, \citet{LaMassa2017} investigated the CL AGN Mrk 1018, which historically transitioned from Type 1.9 to Type 1 between 1979 and 1984 \citep{Cohen1986} before reverting to Type 1.9 with the disappearance of broad emission lines in 2015 \citep{McElroy2016}. This transition was accompanied by the disappearance of the H$\alpha$ line and a concurrent decrease in X-ray flux. Similarly, \citet{Oknyansky2019MNRAS} studied NGC 1566 using multi-wavelength data and found that flux variations in the broad optical lines were strongly correlated with significant changes in the X-ray continuum flux. \citet{Kollatschny2020} investigated IRAS 23226-3843, which transitioned from Type 1 to Type 2 between 1999 and 2017. Their study reported a reduction in broad-line components and optical continuum flux by a factor of 2.5, along with a dramatic decrease in Swift X-ray flux by a factor of 35. Furthermore, \citet{Guolo2021} demonstrated a direct correlation between H$\alpha$ flux and X-ray luminosity for the AGN NGC 2992.

If the CL phenomenon arises from intrinsic changes in the accretion flow properties and is linked to X-ray flux variations, it may also influence the Fe K emission complex. The Fe K line consists of the components:  i) thermal emission from a hot medium, and, ii) fluorescence from a cold medium. Typically, the cold medium dominates the emission, but high-quality data occasionally reveal contributions from highly ionized Fe ions \cite[e.g.,][]{Bianchi2007}. Studies of Fe K-emitting regions, which arise from Fe ions at various ionization levels, indicate that the highly ionized Fe K$\alpha$ component, associated with H-like and He-like ions, most likely originates in the coronal regions located a few tens of gravitational radii from the black hole \citep{Sulentic1998,Ballantyne2002}.

Indeed, if this ionized Fe K$\alpha$ component is produced within the inner accretion flow, its variability may be driven by changes in the coronal X-ray continuum. As the X-ray continuum varies, the Fe K$\alpha$ emission may exhibit corresponding changes. However, due to the limited quality of X-ray spectra, observational studies on the variability of the {highly ionized} Fe K$\alpha$ line during optical/UV CL phenomena remain scarce.

In the case of NGC 2992, \citet{Murphy2007} observed a highly red-shifted Fe K line that coincided with dramatic flaring activity of the $2-10$ keV X-ray flux. \citet{Marinucci2020} further confirmed variability in the Fe K$\alpha$ emission in NGC 2992, correlated with changes in the X-ray continuum flux. For NGC 1365, \citet{Mondal2022} reported that absorption near the Fe K$\alpha$ energy centroid inversely responds to the mass accretion rate: stronger absorption occurs when the accretion rate decreases, and vice versa. Similarly, \citet{Liang2022} studied NGC 1566 and found that an increase in Fe K$\alpha$ flux is in tandem with the 
X-ray flux, supporting the idea that the {highly ionized} Fe K$\alpha$ line originates from disk reflection of coronal X-rays.


In this work, we utilized numerical simulations to  investigate the major BLR emission lines: H$\alpha$ $\lambda6562.80~\AA$, H$\beta$ $\lambda4861.32~\AA$, Mg II $\lambda2798.00~\AA$ and He II $\lambda1640.41~\AA$
and the most prominent X-ray line, Fe K$\alpha$
emission complex, which spans the rest frame energy range of $6-7$ keV, in the AGN system. The primary goal was to explore how these emission properties depend on the broadband continuum generated by the central engine of the AGN. Specifically, we aimed to understand the response of optical/UV emission lines and the ionized Fe K$\alpha$ line to variations in continuum properties, as well as to study the physical conditions of their emitting regions in detail.

To replicate observational results reported in the literature, we analyzed fluctuations in the line luminosities of the selected emission lines in response to changes in the AGN's central continuum emission. The X-ray variations are modeled by adjusting the X-ray to optical/UV slope, $\alpha_{\rm ox}$ (as defined in Eq. \ref{eq:ox}), within the broadband continuum. Moreover, a realistic change in the spectral energy distribution (SED) of the well known CL AGN Mrk 1018 is incorporated, enabling us to study the impact of changing accretion properties on both BLR emission and Fe K$\alpha$ emission. We utilized the latest version of the publicly available photoionization code {\sc cloudy} C23.01 \citep{Gunasekera2023} to simulate a wide grid of parameters and generate the emission line fluxes.

Using the simulated line fluxes, we demonstrated that the optical/UV lines originating at the BLR are weakly sensitive to the changes in the X-ray flux of the continuum emission alone. Moreover, we conclude that the BLR emission, as well as the highly ionized Fe K$\alpha$ line produced at the inner accretion flow, exhibit flux variations in tandem with the changes in the overall shape and the strength of the continuum emission from the AGN, corroborating with the observational results. 

For the case of the CL AGN Mrk 1018, likely triggered by changes in the Eddington ratio, we found that the fluxes of all studied emission lines increase with an increase in the Eddington ratio. The correlation between the Fe K$\alpha$ line emission and continuum variations, alongside BLR emission, supports the hypothesis that the highly ionized Fe K$\alpha$ line originates in the coronal regions of the accretion disk, which undergo intrinsic changes during the CL phenomenon.

The paper is organized as follows. The details regarding the {\sc cloudy} modeling are presented in Section~\ref{sec:cloudy_models}. Section \ref{sec:results} contains the results for the general SED case 
of AGN, while a realistic example of a CL phenomenon in Mrk 1018 is 
discussed in Section~\ref{sec:noda_results}. Finally, the discussion 
of the results and derived conclusions are included in Section~\ref{sec:discussion_conclusions}. Additional figures relevant 
to the results and discussion are shown in the Appendix Section~\ref{sec:appendix}.

\section{Numerical Modeling of the Emission}
\label{sec:cloudy_models}
It is well understood that the presence of spectral features in AGN spectra arise from the reprocessing of continuum radiation emitted by the accretion disk, interacting with gas clouds located at various distances within the system. In this section, we describe the procedure used to model the photo-ionization process and outline the parameters involved in the modeling.
\subsection{Description of the model parameters}
Photo-ionization models are computed for the line emitting regions by
considering the gas clouds irradiated by the central engine 
of the AGN. For this purpose, we utilized the publicly available\footnote{https://gitlab.nublado.org/cloudy/},
numerical simulation code {\sc cloudy} C23.01
\citep{Gunasekera2023}, 
which is an updated version of previous releases of {\sc cloudy} \citep{Ferland2017,Chatzikos2023}. 
The gas clouds are defined to contain \textit{Solar} chemical 
composition adopted from \citet{Grevesse1998} 
\footnote{see hazy1 (\url{https://gitlab.nublado.org/cloudy/cloudy/-/tree/master/docs/latex/hazy1?ref_type=heads}) 
file for the detailed description on the details of the most recent \textit{Solar} composition and its use in {\sc cloudy} C23.01}. The chemical
composition of the AGN environment is not very well constrained albeit several possibilities of metallicity, ranging from the \textit{Solar} to 
a few times \textit{Solar} abundances are proposed \citep{Arav2007,Prez2022}. However, our goal in this paper is to understand how the emission properties of BLR and the Fe K$\alpha$ emitting region change with the CL behavior of the AGN, we adopt the \textit{Solar} chemical composition with an assumption that this choice does not alter the comparison results as long as we assume that the abundances remain constant. 
The Optical/UV and X-ray emission for specific 
lines are modeled using an approximation of a slab of gas illuminated by a continuum SED 
defined in Eq.~\ref{eq:SED_cloudy}.
\begin{equation}
\label{eq:SED_cloudy}
    f_{\nu}=\nu^{\alpha_{\rm uv}} exp\Bigg(\frac{-h\nu}{kT_{BB}}\Bigg)\exp\Bigg( \frac{-kT_{IR}}{h\nu}            \Bigg) +a\nu^{\alpha_{\rm x}}.
\end{equation}
Various components of the AGN spectrum are incorporated in the aforementioned equation. The Big Blue Bump spectrum is
parameterized by the temperature of the bump; $T_{BB}=5.5 \times 10^{5}$ K, and is assumed to have
an infrared exponential cutoff at $kT_{\rm IR} = 0.01$ Rydberg. $\alpha_{\rm uv}$, the low-energy 
slope of the Big Blue Bump is fixed to the standard value $-0.5$. $\alpha_{\rm x}$ represents  
the slope of X-ray 
part, which modifies the X-ray power-law.
We set $\alpha_{\rm x}$ to the value of 
$-0.2$, and the X-ray power-law is extended up to an energy 100 keV.
Beyond this energy, we assume that the photon flux becomes negligible.
These 
parametric values are chosen to represent the standard case for the AGN continuum emission, while also limiting the number of parameters to avoid degeneracies. Nonetheless, parameter values used for comparison, other than those described here, are explicitly detailed where needed. 

The X-ray to UV ratio $\alpha_{\rm ox}$ 
is employed as defined in the Eq.~\ref{eq:ox} \citep{Tananbaum1979,Zamorani1981,Sobolewska2009},
\begin{equation}
\label{eq:ox}
\alpha_{\rm ox}=0.3838~\log\Bigg[\frac{L_{2~keV}}{L_{2500~\AA}}\Bigg],
\end{equation}
where $\alpha_{\rm ox}$ is the spectral index describing the relative
continuum normalization at $2$ keV 
and $2500$ $\AA$. Please note that, we have not used a minus sign in our 
expression of Eq. \ref{eq:ox}. So, the $\alpha_{\rm ox}$ values throughout this 
manuscript will be negative. This convention is adopted to maintain 
a consistency with {\sc cloudy} input which requires the $\alpha_{\rm ox}$ 
values to be provided with their sign explicitly. 
The normalization $a$ in Eq. \ref{eq:SED_cloudy}
is adjusted from the input value of the $\alpha_{\rm ox}$. For the grid of computed models, $\alpha_{\rm ox}$ values in the range $-2.6$ to $-1.0$ are considered.
The choice of this range covers the range of parameter values that are
constrained by AGN surveys in a number of studies \citep{Vagnetti2010,Vagnetti2013}.

\begin{figure}
\centering
\includegraphics[scale=0.61]{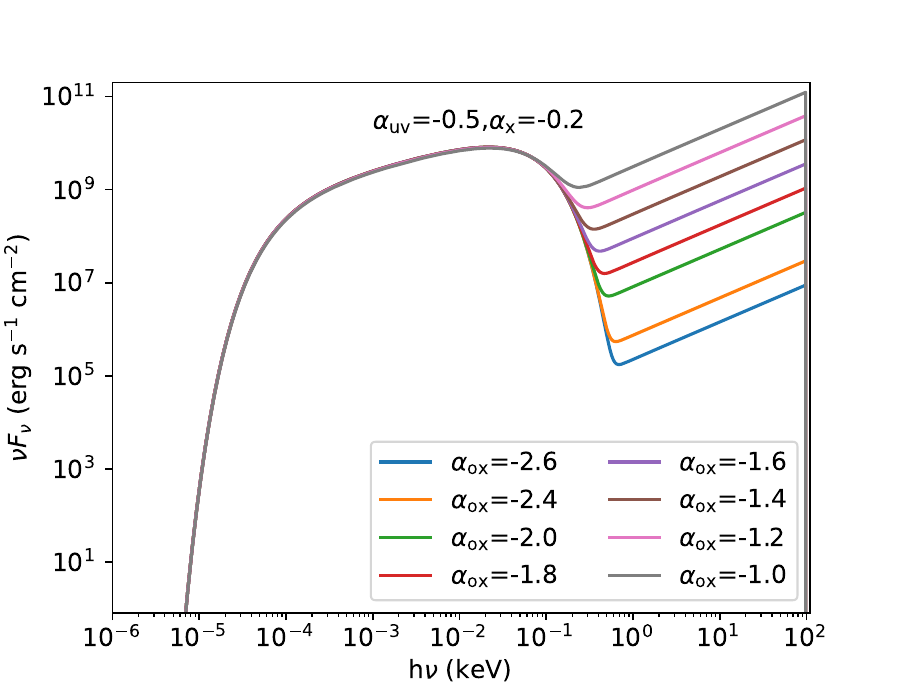}
\caption{AGN continuum radiation fields are shown for $\alpha_{\rm ox}$ values 
in the range $-2.6$ to $-1.0$. The rest of the parameters, $\alpha_{\rm uv}=-0.5$ and $\alpha_{\rm x}=-0.2$, are kept constant.  These SED shapes are used as the incident radiation field in the {\sc cloudy} numerical simulations of the 
line-emitting medium.}
\label{fig:alphaox_vary_SED}
\end{figure}

The resulting SED shapes for varying $\alpha_{\rm ox}$ values 
are presented in Fig.~\ref{fig:alphaox_vary_SED}. 
Note that the SEDs shown in the figure
only provide the shape of the radiation fields. 
The strength of the radiation field is set separately 
by the ionization parameter, defined as:
\begin{equation}
\label{eq:U}
U=\frac{Q(H)}{4\pi r_{0}^2 n_{\rm H} c},
\end{equation}
where $Q(H)$ [s$^{-1}$] is the number of H-ionizing photons, $n_{\rm H}$ is the gas number density, and $r_{0}$ is the inner radius - the distance of the emitting cloud from the source of the radiation field. 

\begin{figure}
\centering
\includegraphics[scale=0.55]{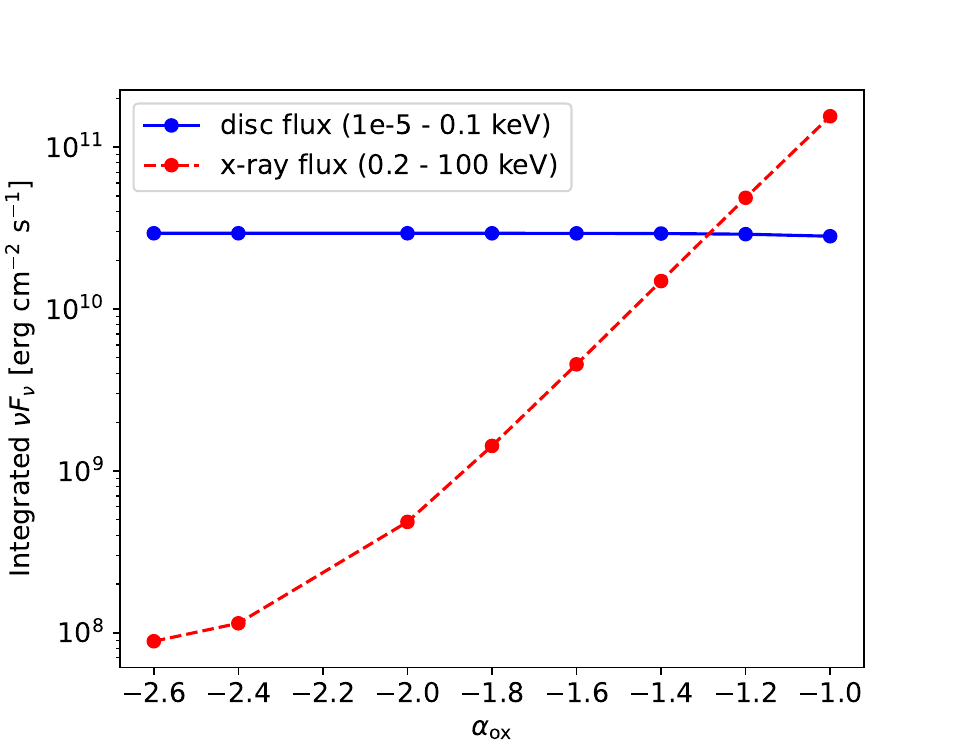}
\caption{Disc flux and X-ray flux of the SEDs presented in Fig~\ref{fig:alphaox_vary_SED}. The disc flux is calculated by integrating the photon fluxes in the energy range $10^{-5} - 0.1$ keV, while the X-ray flux is integrated in the energy range $0.2 - 100$ keV. The fluxes shown are for the ionization parameter value of $\log U = -2.0$ and the density of the medium $n_{\rm H} = 10^{12}$ cm$^{-3}$.}
\label{fig:Xray_intensity}
\end{figure}

Fig.~\ref{fig:Xray_intensity} shows the integrated disc and X-ray fluxes, plotted for the varying values of the $\alpha_{\rm ox}$ parameter, with the SED shapes presented in the Fig. \ref{fig:alphaox_vary_SED}. As expected from a first glance at the SED shapes, the disc flux remains constant, while the X-ray flux rises by $\sim 4$ orders of magnitude with the extreme increase in $\alpha_{\rm ox}$. This plot provides an overall idea of how the total X-ray intensity behaves with the X-ray to UV ratio for a constant disc emission. Please note that the intensity values presented are normalized by n$_{\rm H} =10^{12}$ cm$^{-3}$ and an ionization parameter $U$ = $10^{-2}$ in {\sc cloudy}, and
the normalization factor will vary for other values of the gas 
density and the ionization parameter.

Initially, a large number of models are computed using a grid of
parameter spaces for gas number density and ionization parameter of the emitting medium. This step is quite useful to explore the global effects of various parameters in the estimation of the line fluxes.
Gas density $\log (n_{\rm H}$ [cm$^{-3}$]) ranges from $8-15$ with 
a step size of $1.0$, accommodating the range of BLR gas densities, 
and the ionization parameter 
$\log U$ ranges from $-3.0$ to $5.0$ with a step size of $1.0$. 
We choose the slab thickness, which
defines the material content, by assigning a total column density
of $N_{\rm H} = 10^{23}$ cm$^{-2}$. We note that we also computed models with lower column density values; however, we did not find much difference in the relative nature of the line fluxes, although the numerical values differed. 
Thus, we decided to using a single column density for brevity. A constant density condition with an open geometry is assumed for all simulations. We note that
for the BLR gas, which is relatively dense as compared to the distant
narrow-line region (NLR) gas, there is no significant difference 
between the constant density and constant pressure conditions in photo-ionization simulations. A plane-parallel open geometry, appropriate for describing
the emission regions of the AGN environment, is adopted. We refer to the papers \citet{Adhikari2018,Adhikari2019pmdd,Adhikari2019} for a detailed description of the {\sc cloudy} models in various geometric settings and the differences between constant density and constant pressure conditions applicable to AGN emission and absorption modeling. 
\subsection{Estimation of line strengths}
The most important broad optical/UV lines—H$\alpha$ $\lambda 6562.80 ~\AA$, H$\beta$ $\lambda 4861.32 ~\AA$, Mg II $\lambda 2798.00 ~\AA$, and He II $\lambda 1640.41 ~\AA$—are considered representative emission lines in the optical/UV band for this study. The choice of these lines is motivated by the fact that they are emitted in the BLR, are well-studied using optical/UV observations, and are known to vary during the CL phenomenon in AGN. The line 
centroid wavelengths mentioned are the rest-frame wavelengths as used in the {\sc cloudy} code. 

From past observational studies, during the CL phases of AGN, many of these lines are found to change their strengths considerably.
For simplicity in the modeling process, we assume that the ionization
parameter of the BLR is identical for all the lines considered. 
We note that the real situation might be slightly different, as individual lines may originate from different regions.
Low-ionization lines H$\alpha$, H$\beta$, and Mg II, and high-ionization line He II may originate from different regions with slightly different ionization parameters within the same BLR, as shown by \citet[][and references therein]{Gaskell2009}. This consideration is particularly useful for enabling a direct comparison between the simulated and the observed fluxes, and understanding the covering factor of the emitting region. However, in a broad sense, the differences in ionization parameters responsible for emitting the various broad lines are not significant—particularly in the context of our goal of understanding the relative changes in line fluxes. 
A number of past 
studies have shown that a global description of the BLR parameters
can be used to study the various emission lines originating there \cite[e.g.,][]{Netzer1993,Adhikari2016,Wu2024}.

As our goal in this paper is to compare the emission properties of optical/UV lines and the most notable X-ray emission line, Fe K$\alpha$, we also simulated the flux of the Fe K$\alpha$ line complex centered around 6–7 keV in this study.
The total emission of the Fe K$\alpha$ line can be attributed to the sum of the fluxes from various emission components, produced at different distances within the accretion disk and dusty torus. 
The contributions to the total Fe K$\alpha$ emission from the multiple ionization states of the Fe ions are considered: (i) H-like 1-electron ions; (ii) He-like 2-electron ions; (iii) fluorescent hot emission arising from ions Fe XVIII to Fe XXIII; and (iv) fluorescent cold emission from ions $\leq$ Fe XVII.
We are interested in exploring the changes in the accretion disk properties during the CL phenomenon in an AGN, and therefore, we focus on the highly ionized Fe K$\alpha$ emission originating at the inner radii of the accretion disk close to the black hole. We do not include the dust grains in our simulations, which are mostly involved in studying the cold fluorescence arising from the dusty torus. In the extreme environment of the inner accretion flow close to the black hole, dust grains cannot survive the intense radiation field.
\section{Line fluxes for varying X-ray continuum}
\label{sec:results}
\subsection{Optical/UV line fluxes}
Line fluxes of BLR emission lines: H$\alpha$ $\lambda 6562.80 \AA$, H$\beta$ $\lambda 4861.32 \AA$, Mg II $\lambda 2798.00 \AA$, and He II $\lambda 1640.41 \AA$ are estimated from the 
photo-ionized models of the emitting gas for a large grid of parameters using
the numerical code {\sc cloudy}, and the various aspects of the modeling are
discussed in this section. Firstly, we plotted the line fluxes
as a function of the 3-dimensional (3D) parameter spaces in $n_{\rm H}$, $\alpha_{\rm ox}$ 
and $U$, looking for the flux dependence on these parameters. 
This step is very useful to reduce the number of parameters and 
select the most appropriate ones for the further investigation.

\begin{figure}
\centering
     \includegraphics[scale=0.7]{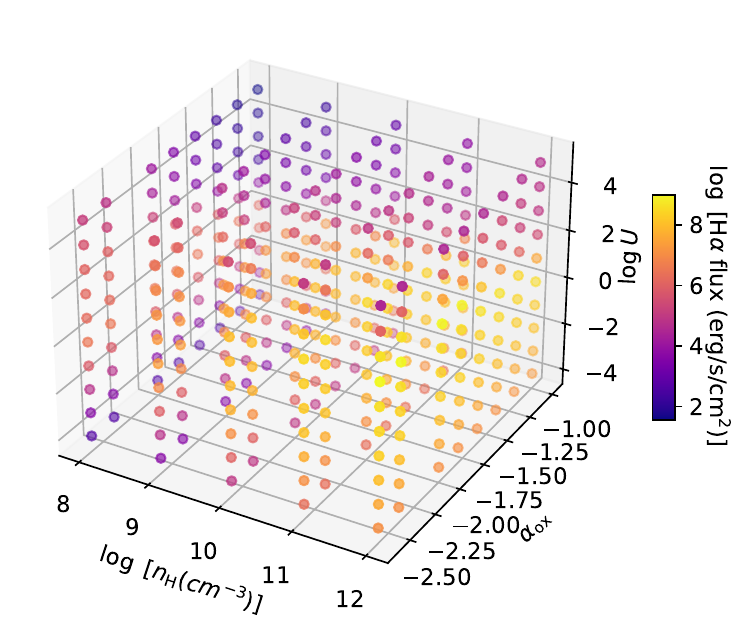}
    \caption{H$\alpha$ emission plotted for a 3D parameter space: gas density $n_{\rm H}$, X-ray to optical slope $\alpha_{\rm ox}$, and the ionization parameter $U$. The BLR clouds are illuminated by the SEDs shown in Fig. \ref{fig:alphaox_vary_SED}.}
    \label{fig:optical3D_Ha_A}
\end{figure}

\begin{figure}
    \includegraphics[scale=0.7]{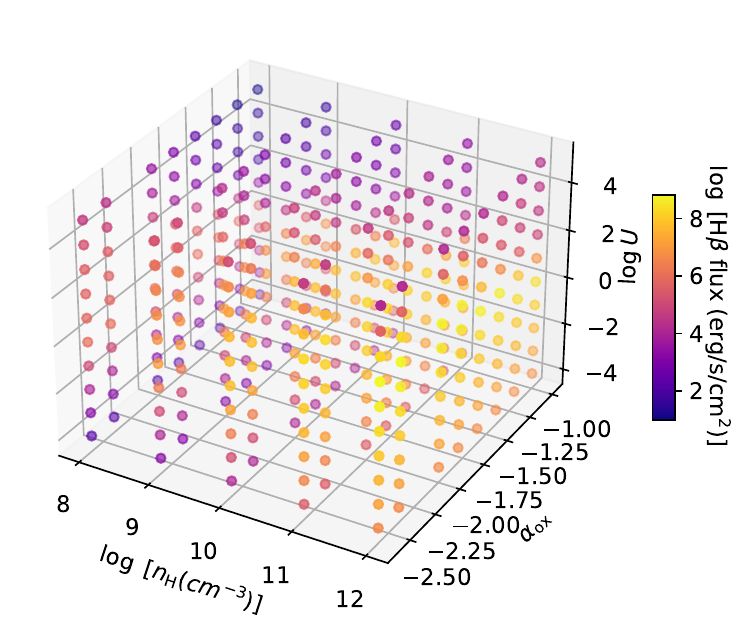}
    \caption{H$\beta$ emission plotted for a 3D parameter space same as described in Fig.\ref{fig:optical3D_Ha_A}.}
    \label{fig:optical3D_Hb_A}
\end{figure}

\begin{figure}
\centering
    \includegraphics[scale=0.7]{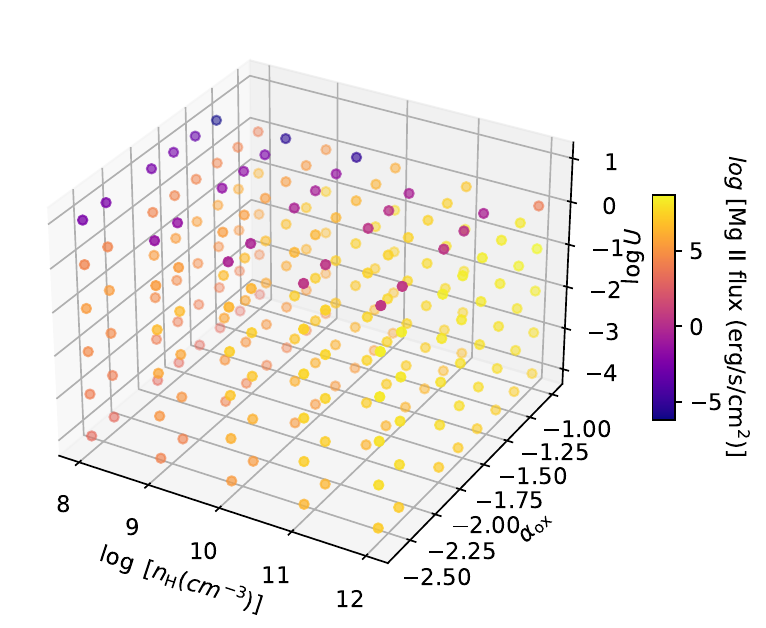}
    \caption{Mg II emission plotted for a 3D parameter space same as described in Fig.\ref{fig:optical3D_Ha_A}. }
    \label{fig:optical3D_Mgii_A}
\end{figure}

\begin{figure}
\centering
    \includegraphics[scale=0.7]{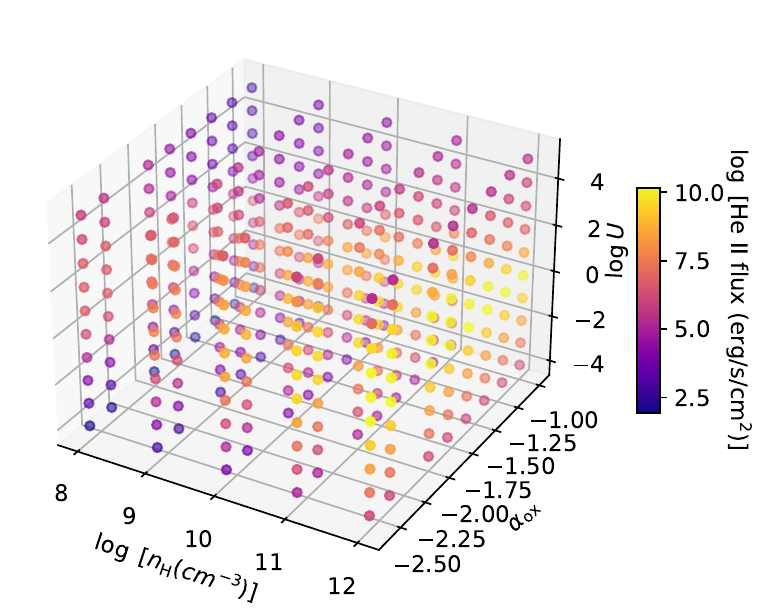}
    \caption{He II emission plotted for a 3D parameter space same as described in Fig.\ref{fig:optical3D_Ha_A}.}
    \label{fig:optical3D_Heii_A}
\end{figure}

We present the 3D plots of the H$\alpha$, H$\beta$, Mg II, and He II
emission strengths in Figs. \ref{fig:optical3D_Ha_A}, \ref{fig:optical3D_Hb_A}, 
\ref{fig:optical3D_Mgii_A}, and \ref{fig:optical3D_Heii_A} respectively. 
The models of emission are calculated for the SEDs displayed in the 
Fig. \ref{fig:alphaox_vary_SED}, whose disc emission and X-ray emission flux behave with the $\alpha_{\rm ox}$ values as shown in Fig. \ref{fig:Xray_intensity}. As seen in the Fig. \ref{fig:optical3D_Ha_A} and \ref{fig:optical3D_Hb_A},  
the flux of all the BLR emission lines clearly depends on the density of 
the emitting medium. The line strengths exhibit an increasing trend with the 
increase in the gas number density of the emitting gas clouds. 
This density dependence of line emission is rather smaller in the case 
of the low-ionization line Mg II which can be seen in Fig. \ref{fig:optical3D_Mgii_A}.
However, the fluxes of all the BLR lines tend to saturate beyond a 
gas density of $10^{11}$ cm$^{-3}$. We did not find a significant difference 
between the fluxes computed for the models with densities $10^{11}$ and $10^{12}$ cm$^{-3}$.  Originally, we simulated the models with higher densities up to $10^{15}$ cm$^{-3}$, but we encountered several convergence issues in {\sc cloudy} and therefore refrain from including these results here.

Moreover, a strong dependence of the line emission on the ionization parameter
is present in the case of all the considered BLR emission lines. Generally, 
the line's flux reaches its maximum at $\log U = -1.0$ for all the lines.
Nevertheless, this dependence is a bit complex as the line flux can peak at 
slightly different values of the ionization parameter for different gas
densities. It is very difficult to distinguish between the nature of variation
of the  H$\alpha$ and H$\beta$ flux with the ionization parameter by 
comparing the 3D plots shown in the Figs. \ref{fig:optical3D_Ha_A} and
\ref{fig:optical3D_Hb_A}. These plots look almost identical with each other. 
Both emission lines peak in the region with the ionization 
parameter $\sim \log U = -1.0$, and become weaker with the 
increase or decrease in the ionization strength. 
As shown in Fig. \ref{fig:optical3D_Mgii_A}, Mg II flux becomes 
strongest around $\log U \sim -1.0$, and with the increase in the 
ionization parameter, it becomes insignificant and is not 
visible. He II flux,
as shown in Fig. \ref{fig:optical3D_Heii_A}, is the most 
significant at $\log U \sim -1.0$, $\sim$ two orders 
of magnitude higher in emission as compared to the peak 
emission of the other lines. 

\begin{figure}
    \centering
    \includegraphics[scale=0.55]{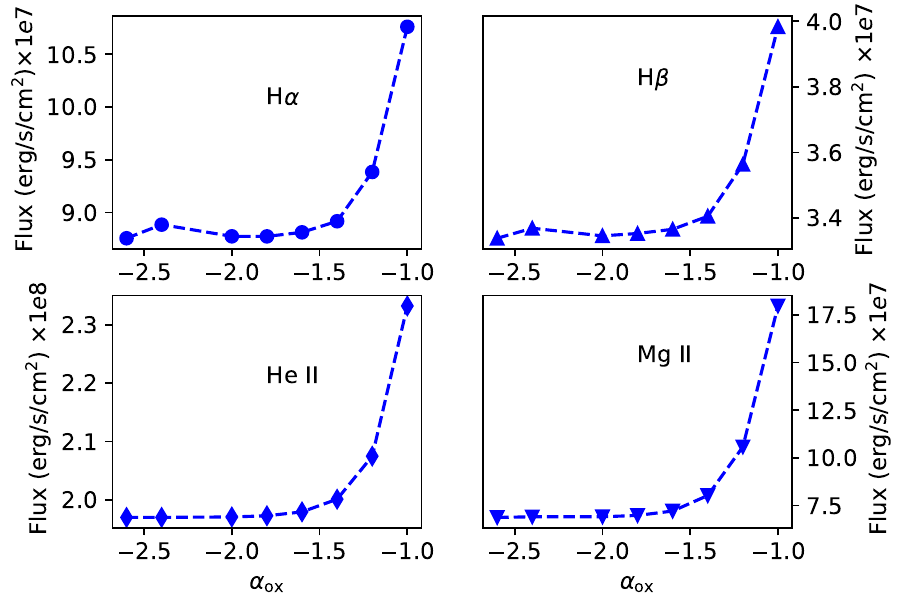}
    \caption{Line emission fluxes for all the considered optical/ UV lines,
    computed by assuming a standard BLR model with $n_{\rm H} = 10^{12}$ cm$^{-3}$, and $\log U = -2.0$, irradiated with the SEDs shown in Fig. \ref{fig:alphaox_vary_SED}. Each panel in the figure corresponds to a specific emission line, as labeled.}
    \label{fig:Lines_optical} 
\end{figure}

The 3D figures also depict that there is a rather weak dependence of optical/UV emission on the $\alpha_{\rm ox}$. However, as the difference in color coding corresponds to the orders of magnitude variation in the line strength, it is hard to notice the flux variations by a few factors. Nevertheless,
the slightly increasing line fluxes for the increasing value of $\alpha_{\rm ox}$ can be noticed in all the 3D plots. This variation is further explored in the Fig. \ref{fig:Lines_optical}, where a dependence of line fluxes on the $\alpha_{\rm ox}$ parameter is shown.

In the next step, we discuss the standard model of BLR emission for a set of 
parameters: $n_{\rm H}$= $10^{12}$ cm$^{-3}$ and $U$= $10^{-2}$.
The choice of these  parameters of BLR region is based on the proposition
that the emitting clouds emerge from the accretion disk
atmosphere\citep{Czerny2011}. With this consideration, \citet{Adhikari2016} have
solved the vertical structure of the standard accretion disk \citep{Shakura1973}
with the assumption of the gray gas opacities \citep{Rozanska1999} and
presented the radial density profile of the disk atmosphere. 
The values, $n_{\rm H}$ = $10^{12}$ and $U$ = $10^{-2}$, adopted here, 
are inside the dust sublimation radius in their density profile 
(see their Figs.9 and 10), and resemble to the BLR region. These BLR parameters also resemble with the values constrained by \citet{Hryniewicz2012} for  the quasar SDSS J094533.99+100950.1. Moreover, 
the universal values of the BLR parameters, the 
density $n_{\rm H} = 10^{12}$ cm$^{-3}$ and the ionization 
parameter $U$= $10^{-2}$ have been shown to qualitatively 
reproduce the observed radius-luminosity relation for many AGN by
\citet{Wu2024}. Motivated by aforementioned results, we further 
discuss the BLR emission based on the models computed with 
density and ionization parameters fixed at these values.

Fig. \ref{fig:Lines_optical} shows the variation in the emission of all four BLR lines as a function of the $\alpha_{\rm ox}$ values. As seen in Fig. \ref{fig:Xray_intensity}, the disc flux in our incident SEDs remains constant while the X-ray power-law flux significantly increases with the rise in $\alpha_{\rm ox}$. For the increasing X-ray intensity of the incident radiation field up to the region where $\alpha_{\rm ox} \leq -1.6$, no change in the emission line fluxes is observed.
However, with the further increase in the incident X-ray intensity, the emission line fluxes slightly increase. Beyond this point, $\alpha_{\rm ox} > -1.6$, when the X-ray intensity of the radiation field varies by $\sim 2$ orders of magnitude, the H$\alpha$ and H$\beta$ emission flux increases by a small factor, with the least increase for H$\beta$ (by a factor of $\sim 1.2$) and the largest for H$\alpha$ (by a factor of $\sim 1.3$). For Mg II, the increase in flux is by a slightly higher factor, $\sim 2.4$, compared to the H$\alpha$ and H$\beta$ lines. The case of He II is similar, with the emission flux increasing only by a factor of $\sim 1.2$.

To account for the fact that the X-ray power law slope can vary with the Eddington ratio \citep{Risaliti2009,Gupta2024}, we also simulated models using standard BLR parameters for softer X-ray slopes ($\alpha_{\rm x}=-0.5$ and $-1.0$). A comparison of the resulting line fluxes is shown in Fig.~\ref{fig:optical_flux_alphaX},
in section \ref{sec:appendix}. We did not find any noticeable difference in the trend of variations across the different X-ray slopes in the SED. 

Moreover, we examined line flux variations at lower disk temperatures ($T_{\rm BB}=5\times10^{4}$ and $10^{5}$ K) and present the results in Fig.~\ref{fig:optical_flux_var_T},
in the section~\ref{sec:appendix}. We found a noticeable difference in the flux levels: H${\alpha}$, H$\beta$ and Mg II line fluxes are higher at the lowest disk temperature, while the He II flux is higher at the highest disk temperature. This is as expected, as H${\alpha}$, H$\beta$ and Mg II are low-ionization lines, whereas He II is a high ionization line. Nevertheless, the relative flux variation with changing X-ray illumination follows a similar trend across temperatures, except for the He II line. At the lowest $T_{BB}$, the He II flux increases rapidly under intense X-ray illumination. 

These results show that if only the X-ray flux of the SED illuminating the BLR region changes while the disc flux remains constant, the optical/UV broad line emission is not significantly altered. Only an enhancement by a small factor in the line fluxes is observed. The optical/UV line emission is more sensitive to the disc emission, which remains largely unchanged in the SEDs we have considered. This result highlights the need for changes in the disc emission during the CL phenomenon of AGNs, especially in cases where BLR emission lines clearly appear or disappear. Moreover, the requirement of substantial X-ray flux variations of the incident continuum to produce subtle changes in the optical/UV emission flux in these models does not align with observational results, indicating that some process occurring in the innermost accretion disk is at play in such types of CL AGNs.

\subsection{Fe K$\alpha$ emission}
The contribution to the Fe K$\alpha$ emission from the
medium close to the accretion disk of AGN is simulated using {\sc cloudy}. 
As of now, it is quite well known that Fe K emission is produced across the
different scales of the accretion disk \citep{Sulentic1998,Marinucci2018,Marinucci2020}, with a highly ionized component from the
close environment of the black hole \citep{Fabian1989,Iwasawa2012} and a narrow
component originating from the reprocessing of the X-rays by the dusty
torus\citep{Yaqoob2004,Shu2010,Gohil2015}.
Several new studies have also reported that the narrow component of the Fe K$\alpha$ line
originates over a range of distances, from  the BLR extending out to the dusty torus\citep{Gandhi2015,Andonie2022,Xrism2024ApJ}.
To systematically 
study the dependence of the Fe K$\alpha$ flux on the grid of model parameters,
we first calculated the contributions of the various emission components to
the total Fe K$\alpha$ flux. We are aware with the fact that {\sc cloudy}
calculation of the Fe K flux is limited by the assumption of stationary
gas. Thus, the dynamics and general relativistic effects, which play a vital role in broadening the line, are not taken into account here. However, for our purpose of understanding the link between continuum variation, UV/optical emission, and Fe K emission during the CL phenomenon of AGN, the {\sc cloudy} estimation and comparison of the line fluxes serve this purpose.

\begin{figure}
    \includegraphics[scale=0.7]{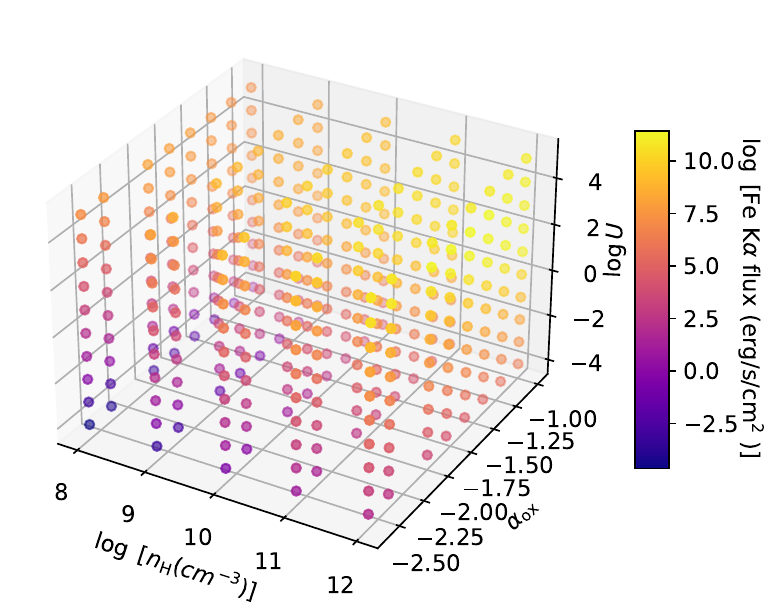}
    \caption{Total Fe K$\alpha$ flux: the sum of the cold and hot Fe
    emission components: (i) H-like ions, (ii) He -like ions,
    (iii) hot ions (Fe XVIII - Fe XXIII), and (iv) cold ions ($\leq$ Fe XVII),
    plotted as a function of  3D parameter space: gas density $n_{\rm H}$, 
    X-ray to optical slope $\alpha_{\rm ox}$, and ionization strength $U$.}
    \label{fig:Fe_k_alpha_3D_A}
\end{figure}

\begin{figure}
    \includegraphics[scale=0.7]{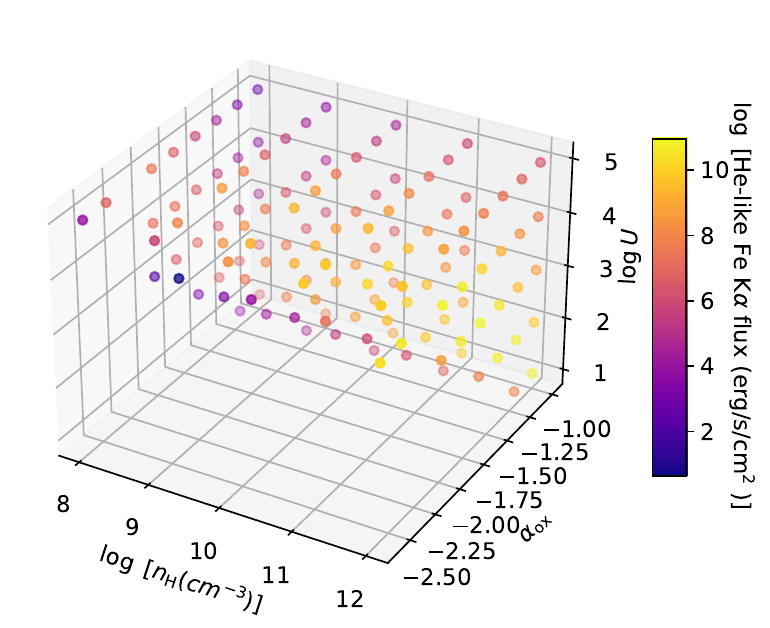}
    \caption{3D plot of emission due to He-like ions of Fe atom.}
    \label{fig:Fe_He_3D}
\end{figure}
\begin{figure}
    \includegraphics[scale=0.7]{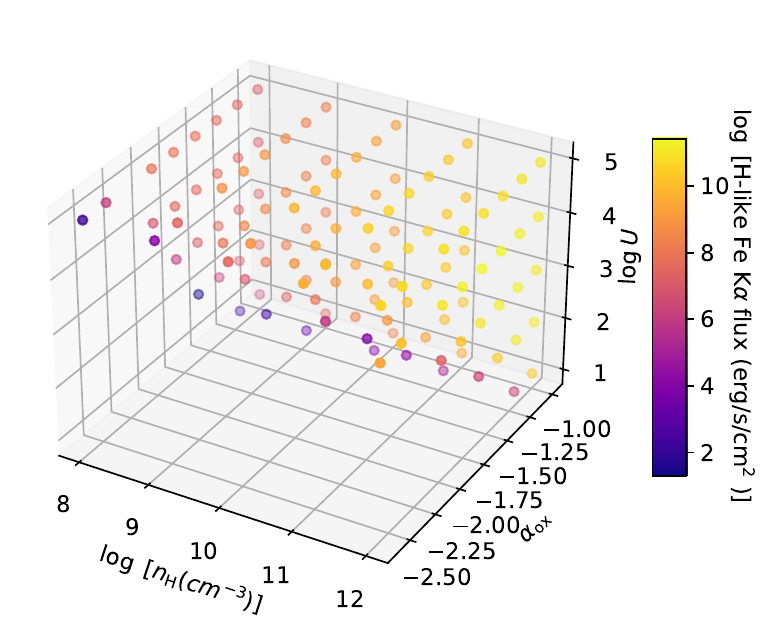}
    \caption{3D plot of emission due to H-like ions of Fe atom.}
    \label{fig:Fe_H_3D}
\end{figure}

Fig. \ref{fig:Fe_k_alpha_3D_A} depicts a 3D plot of the total Fe K$\alpha$ flux, computed for the models of the emitting region irradiated with the SEDs shown in Fig. \ref{fig:alphaox_vary_SED}. This Fe K flux includes the sum of all the contributions: hot and cold Fe emission around $6-7$ keV. The emission flux is clearly dependent on $n_{\rm H}$, $U$, and $\alpha_{\rm ox}$. As shown in the 3D plot, pronounced Fe K$\alpha$ emission clusters in regions with high gas density and high ionization parameter. There is also a clear trend of increasing total Fe K emission as $\alpha_{\rm ox}$ (X-ray illumination) increases. This is quite obvious, as more incident high-energy X-ray photons result in a higher flux of Fe K emission.

As the ionization parameter decreases from $\log U=0$, the total emission weakens. In regions modeled with high ionization degrees, the total emission is primarily dominated by H-like and He-like Fe ions. However, at lower ionization parameters, the populations of H-like and He-like ions drop significantly, making their emission negligible and invisible. A similar trend of negligible H-like and He-like Fe K emission is observed in models with low X-ray irradiation (SEDs with lower $\alpha_{\rm ox}$ values). These features are evident in the 3D plots of He-like and H-like emission flux shown in Figs. \ref{fig:Fe_He_3D} and \ref{fig:Fe_H_3D}, respectively.
\begin{figure}
    \includegraphics[scale=0.7]{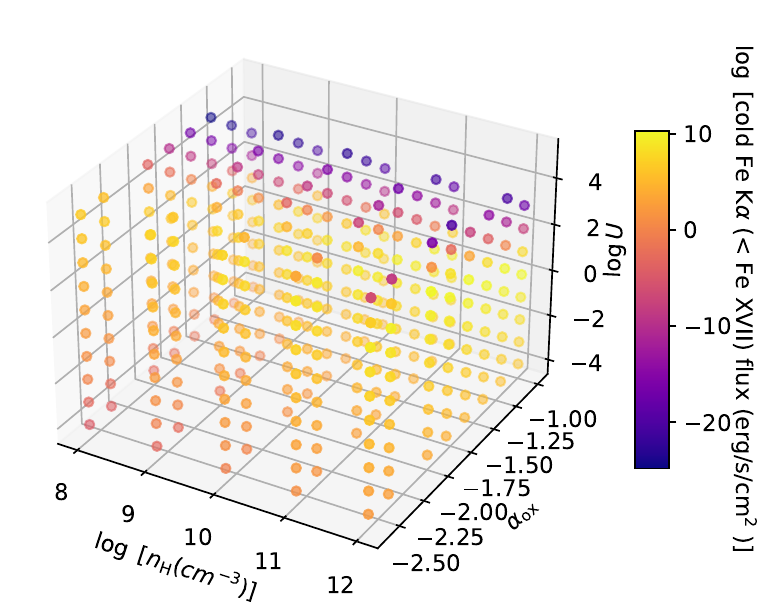}
    \caption{3D plot of fluorescent Fe K$\alpha$ emission due to 
    cold Fe ions with ionization levels $\leq$ Fe XVII.}
    \label{fig:Fe_cold_3D}
\end{figure}
\begin{figure}
    \includegraphics[scale=0.7]{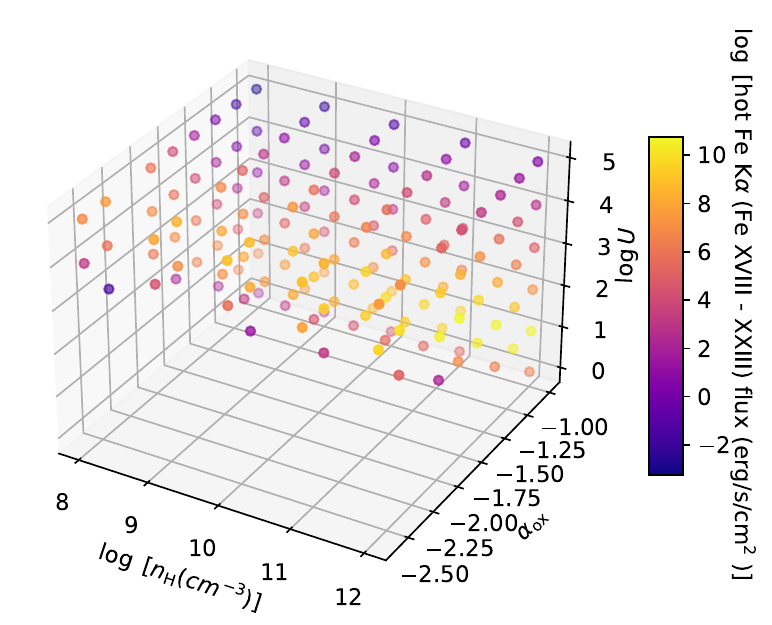}
    \caption{3D plot of fluorescence Fe K$\alpha$ emission due to 
    the hot ions (Fe XVIII - Fe XXIII).}
    \label{fig:Fe_hot_3D}
\end{figure}
The cold fluorescence emission contributed by Fe ions with levels $\leq$ Fe XVII and the hot fluorescence flux from ions with levels in the range Fe XVIII - Fe XXIII are shown in the Figs. \ref{fig:Fe_cold_3D} and \ref{fig:Fe_hot_3D} respectively. 
As expected, the cold fluorescence emission is mostly associated with low-ionization parameter models, while the hot fluorescence emission primarily originates from regions with moderate ionization parameters. Therefore, the fluorescence emission at the inner scales of the accretion disk is insignificant. These results clearly indicate that, 
the production of a significant amount of highly ionized Fe K flux is associated with the region close to the source of irradiation, i.e., the region of interest for this study. Our result aligns with several findings in the literature, which similarly conclude that a significant portion of the 
highly ionized Fe K$\alpha$ emission originates from the reflection of the primary X-ray photons in the inner environment of the accretion disk \citep{Sulentic1998,Yaqoob2003,Reeves2001}.

In the next step, and in line with these findings, we consider a model of highly ionized gas that predominantly produces H-like and He-like Fe K emission. Thus, we deliberately minimize the contribution of fluorescence emission to focus on quantifying the highly ionized Fe K flux and studying its dependence on variations in X-ray illumination. This choice also justifies excluding dust grains in the {\sc cloudy} simulations, as they can not survive the extreme environment where
the highly ionized Fe K$\alpha$ is originating. To estimate the Fe K flux, we adopt a model with a gas density of $\log[n_{\rm H}] = 12$ and $\log[U] = 3.0$. This density choice remains consistent with the radial density profile computed by \citet{Adhikari2016} for the atmosphere of the standard accretion disk. The higher ionization parameter, compared to that of the BLR, accounts for the fact that the highly ionized Fe K$\alpha$ emitting region is exposed to a more intense radiation field, being closer to the X-ray source. This consideration of the high ionization parameter is also consistent with values derived for ultra-fast outflows (UFOs) originating from the inner regions of the disk, observed through Fe XXV and Fe XXVI K-shell absorption lines in the hard X-ray band \citep[see][for a review on outflows in AGN]{Laha2021}.

\begin{figure}
    \centering
    \includegraphics[scale=0.6]{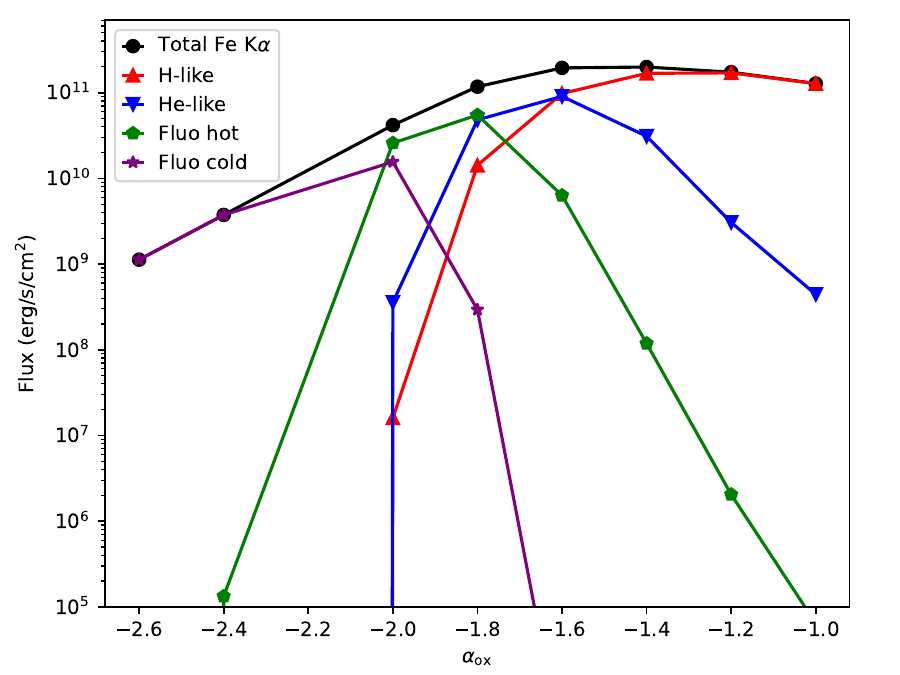}
    \caption{Contributions to the total Fe K flux estimated from a model of highly ionized gas located close to the 
    source of irradiation, defined by the parameters $n_{\rm H} = 10^{12}$ cm$^{-3}$ and $\log U = 3.0$. The total Fe K$\alpha$ flux (black points)
    represents the sum of emission from various Fe ions : H-like (red upward-facing triangles), He-like (blue downward-facing triangles), fluorescent hot (Fe XVIII–XXIII) (green pentagons), and fluorescent cold ($\leq$ Fe XVII) (purple stars).}
    \label{fig:Lines_Fe} 
\end{figure}

The variation of Fe K$\alpha$ flux, predominantly contributed by the 
emission from H-like and He-like ions of Fe, is plotted against the changing 
strength of X-ray irradiation, expressed in terms of $\alpha_{\rm ox}$, in Fig. \ref{fig:Lines_Fe}. As observed from the plot, with an increase in the X-ray illumination, the total Fe K$\alpha$ flux (black points with line) rises sharply and attains a peak value. With extreme variations in the incident X-ray flux of the SED by $\sim 3$ orders of magnitude, the Fe K flux increases by $\sim 2$ orders of magnitude.
Furthermore, beyond the $\alpha_{\rm ox}$ value of $-1.6$, the Fe K$\alpha$ emission plateaus. For the extreme SEDs with high-intensity X-ray irradiation beyond $\alpha_{\rm ox} = -1.2$, a declining trend in emission is noticeable. This saturation in line flux occurs due to the full ionization of Fe ions. Once the Fe atoms become fully ionized, the medium becomes transparent to the increasing energy of the hard X-ray photons. This behavior of Fe K$\alpha$ flux with varying X-ray irradiation corroborates with the findings of \citet{Liang2022}, where the authors reported that, in the case of the CL AGN NGC 1566, the Fe K$\alpha$ flux responded in tandem to a substantial increase in the hard X-ray continuum flux.

The contribution to the total Fe K$\alpha$ emission from the various ionization states of the Fe atom is also shown in Fig. \ref{fig:Lines_Fe}.
Initially, for the lowest X-ray intensity, the total emission is dominated by the fluorescent cold emission from the Fe ions $\leq$ Fe XVII. 
This result is in excellent agreement with the conclusion of \citet{Ballantyne2002}, which demonstrated that for low X-ray irradiation, the Fe K emission is mostly due to cold fluorescence. However, with the rise in the X-ray intensity of the incident radiation field, the cold emission quickly drops and the fluorescent hot emission prevails as expected due to the increase in the population of FeXVIII - FeXXIII ions. With a further increase in the X-ray intensity, fluorescent hot emission decreases significantly, and the total emission is predominantly from the H-like and He-like Fe ions. 
Further increase in the X-ray irradiation beyond $\alpha_{\rm ox} = -1.4$ results in a rapid drop in He-like emission as Fe becomes fully ionized, and the emission is solely from H-like ions. A similar observation of enhanced Fe K emission due to H-like ions in the case of an X-ray illuminated accretion disk of X-ray binaries was noted by \citet{Mondal2021}.

The photo-ionization models of the BLR and the Fe K line region developed in this section demonstrate that the optical/UV BLR lines are weakly sensitive to changes in the X-ray flux of the irradiating continuum. However, this result does not sufficiently explain the observed correlation between the disappearance/appearance of broad emission lines and the decrease/increase in the X-ray continuum flux. On the other hand, we find that the highly ionized Fe K$\alpha$ flux strongly depends on the strength of the X-ray irradiation. Nevertheless, the SED used in these models may differ from realistic changes in the ionizing radiation fields, which often occur simultaneously with the appearance/disappearance of the broad emission lines. Furthermore, there is growing consensus among experts that a true CL phenomenon is likely related to changes in the intrinsic properties of the accretion disk. The change in the accretion rate, among other possibilities, has been proposed as a main mechanism driving the CL phenomenon in AGNs, as suggested in a number of recent studies \citep{Noda2018,Yang2023,Veronese2024}. If this is the case, it is valuable to consider the realistic nature of SED variations across a broad range of energies and to compute emission models for estimating the line fluxes. Such investigation will provide quantitative information on how the emission line fluxes considered in this manuscript respond to realistic changes in the irradiating SED during a CL phenomenon. In the following section \ref{sec:noda_results}, we describe such considerations in detail.
\section{Line fluxes for varying broad band continuum}
\label{sec:noda_results}

In this section, we studied the line emission for the realistic case of a CL AGN by considering the fact that this phenomenon may involve changes in both the shape and flux of the broadband continuum radiation field. For this consideration, it is crucial to include the effects of variations in the optical/UV part of the AGN emission, which primarily originates from the accretion disk, as well as changes in the X-ray power-law component.

A good example of such a realistic case is the AGN Mrk 1018, which has displayed repeated spectral type changes throughout its history of observations \citep{Cohen1986,McElroy2016}. \citet[][hereafter ND18]{Noda2018} investigated its transition from Sy 1 to Sy 1.9 over an eight-year time span, examining its CL properties in detail. ND18 extensively studied this source using multi-wavelength observations conducted between 2008 and 2016 and demonstrated that the entire broadband spectrum undergoes significant changes during its transition from the brightest to the faintest phase.
Notably, the authors showed that the disk emission, the associated soft X-ray excess, and the X-ray power-law component undergo dramatic variations during this period.
ND18 explained this state transition as being driven by intrinsic changes in the accretion flow properties, transitioning from a bright accretion disk to a hot inner flow characteristic of the advection-dominated accretion phase.
Very recently, \citet{Veronese2024} proposed a possible mechanism triggering
this transition in Mrk 1018 by speculating that gaseous clouds are pushed onto the innermost regions of the AGN by
a galactic or an extragalactic process causing the accretion flow to puff up to the advection dominated flow.
This result is particularly interesting because it provides a potential explanation for the driving mechanism behind CL properties in AGNs in general.

In our work, we utilized the true shapes and luminosities of the broadband continuum SEDs of Mrk 1018, as constrained by ND18, and simulated the line-emitting regions using {\sc cloudy}. We then analyzed the resulting behavior of line fluxes and their dependence on the varying shapes and strengths of the continuum radiation, corresponding to the state transitions of Mrk 1018 as proposed by ND18. Although we use Mrk 1018 as a specific example of a CL AGN, 
our overall goal is to explore the general case of AGN undergoing similar state changes during CL phenomena. 

\begin{figure}
    \centering
    \includegraphics[scale=0.6]{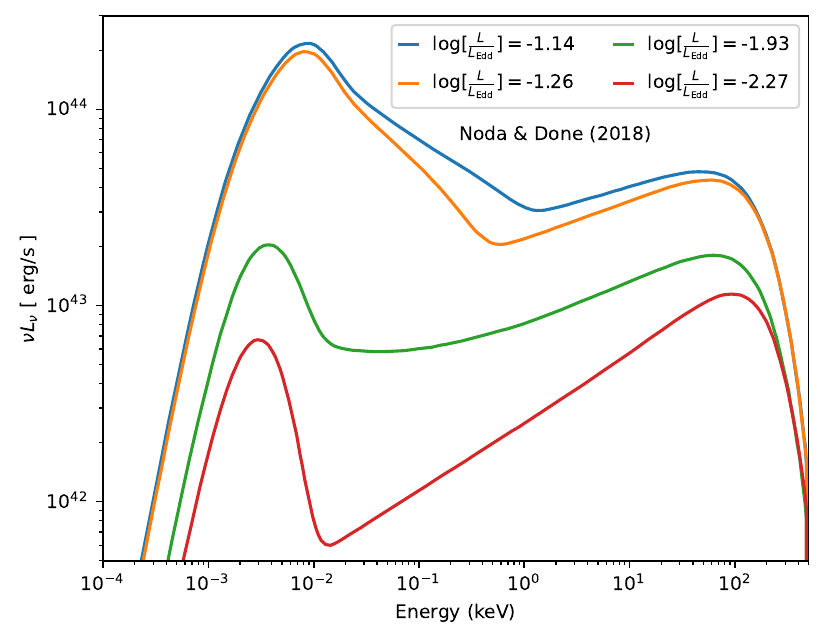}
    \caption{The SEDs of Mrk 1018, taken from \citet{Noda2018}
    and normalized by the bolometric luminosities. These SEDs 
    are used as an incident radiation field for 
    the {\sc cloudy} models presented in Section~\ref{sec:noda_results}.}
    \label{fig:sed_noda}
\end{figure}

\begin{figure}
    \centering
    \includegraphics[scale=0.6]{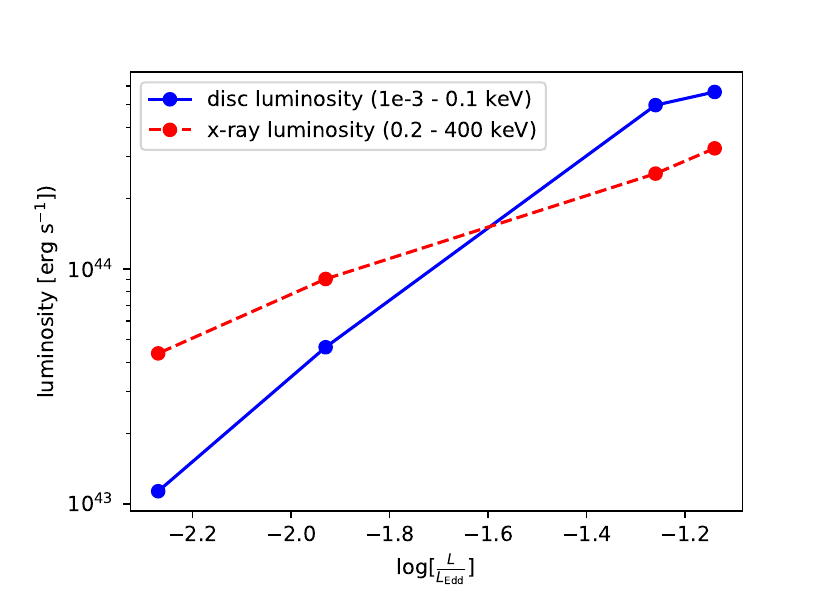}
    \caption{The disc and X-ray power-law luminosities during the CL phenomenon of Mrk 1018, obtained by integrating the continuum luminosities over the defined energy bands versus Eddington ratio is plotted.}
    \label{fig:both_intensity_noda}
\end{figure}

\begin{figure}
    \centering
    \includegraphics[scale=0.6]{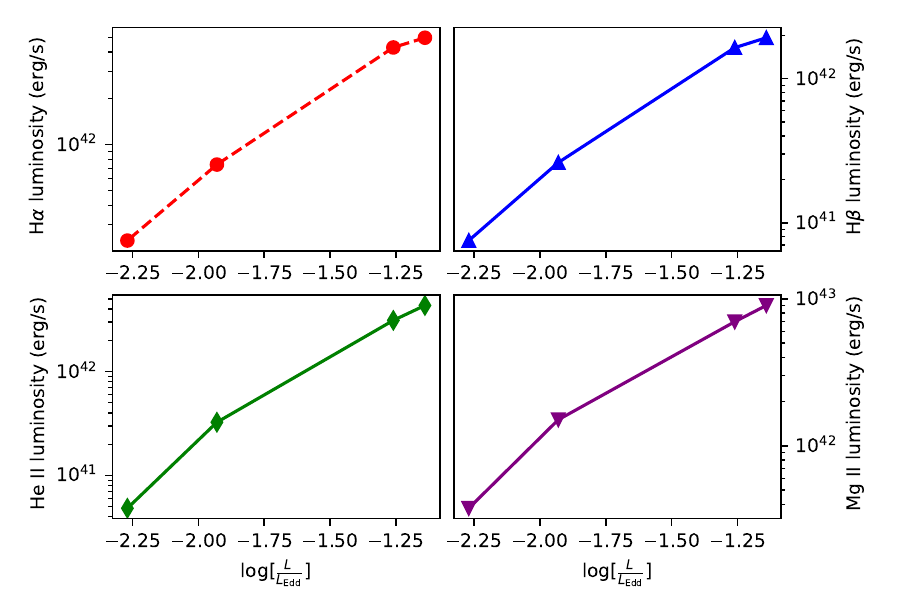}
    \caption{BLR line luminosities plotted as a function of the change in strength and shape of the irradiation fields, expressed in terms of the varying Eddington ratio $\frac{L}{L_{\rm Edd}}$ of Mrk 1018.}
    \label{fig:flux_noda}
\end{figure}

Fig. \ref{fig:sed_noda}, reproduced from ND18 (see their Fig. 1, right panel), shows the changes in the shape and strength of the continuum emission during the CL phenomenon of Mrk 1018 over the 8 years of observations. 
It is clear from the figure that dramatic changes in the type of accretion occurred during this phase. Furthermore, we divided the total continuum emission into contributions from two components: (i) the disc part, obtained by integrating the luminosities from the lowest energy up to 0.1 keV, and (ii) the X-ray power-law part, integrated in the energy range 0.2-400 keV. The change in luminosity in these two components versus $\frac {L}{L_{\rm Edd}}$ ratio during the CL states of Mrk 1018, is shown in Fig. \ref{fig:both_intensity_noda}. Please note that the mass of the black hole is fixed at the value of $6.92\times 10^{7}M_{\odot}$ by ND18 in the spectral fitting process. Mrk 1018 shows significant variations in both defined continuum bands: a decreasing trend of emission as the AGN changes its type from Sy 1 to Sy 1.9. It is clear from the plot that the X-ray luminosity decreases by $\sim 2$ orders of magnitude, while the disc emission lowers by $\sim$ an order of magnitude.
\subsection{BLR emission models}
To set up the {\sc cloudy} models of the emission region, suitable for this transition from the standard accretion to the advection-dominated accretion flow, we need to calculate the physical radius of the BLR consistent with the change in luminosity of the source. For this purpose, we calculated the BLR radius $R_{\rm BLR}$, which changes with the luminosity of the source, first established by \citet{Kaspi2000}. We use the updated version of their relation, as modified by \citet{Bentz2009}, given in Eq. \ref{eq:Bentz}: 
\begin{equation}
\log R_{\rm BLR}[{\rm H}\beta] = 1.538 \pm 0.027 + 0.5 \log L_{44, 5100}, \label{eq:Bentz} 
\end{equation}
where $R_{\rm BLR}$ is the radius of the BLR in light-days, and $L_{44,5100}$ is the luminosity at $5100\AA$, normalized by the value of $10^{44}$ erg s$^{-1}$. We calculated the bolometric luminosities $L_{\rm bol}$ for each phase of the transition by integrating the associated SEDs shown in Fig. \ref{fig:sed_noda}. During the CL phase of Mrk 1018, the bolometric luminosity drops by a factor of 16, from its brightest to the faintest state. Corresponding to this, ND18 estimated the Eddington ratio $\log \frac{L}{L_{\rm Edd}}$ for each state, which drops from the highest value of $-1.14$ to the lowest value of $-2.27$ (for a black hole mass fixed at the value of $6.92\times 10^{7}M_{\odot}$).

Further, we assumed that the BLR gas density is $n_{\rm H} = 12$ cm$^{-3}$, the standard value adopted in the previous section \ref{sec:results}. For the given value of $n_{\rm H}$, the ionization parameter $U$ is self-consistently calculated implicitly in {\sc cloudy} using Eq. \ref{eq:U}. These models of emitting regions are now more realistic in the sense that we have successfully incorporated the changes in the SED shape and flux, as well as the resulting location of the emitting regions and ionization parameter. The obtained attributes: $\frac{L}{L_{\rm EDD}}$, $L_{\rm bol}$, $L_{5100}$, and $U$ for all four states of the source are summarized in Table \ref{tab:observation}.
\begin{table}
\scriptsize
\centering
\caption{\label{tab:observation} The change in the properties of Mrk 1018
during its transition from Sy1 to Sy1.9 over the 8 years of observations.
From the first to the fifth column, we list: the Eddington ratio, bolometric luminosity, luminosity at $5100 \AA$, radius of the BLR, and the ionization parameter, respectively. The tabulated $\log \frac{L}{L_{\rm Edd}}$ are taken from ND18 and the $L_{\rm bol}$ are calculated by integrating the SEDs presented in Fig. \ref{fig:sed_noda}. $L_{5100}$ is also determined from the SEDs in Fig. \ref{fig:sed_noda}.}
\begin{tabular}{ccccccc}
\hline
  $\log \frac{L}{L_{\rm Edd}}$ &$L_{\rm bol}$ &$L_{5100}$ &$R_{\rm BLR}$ &U&\\
       &    (erg s$^{-1}$)   & (erg s$^{-1}$) &(parsec)&\\
\hline
-1.14&$9.01\times10^{44}$ &$8.88\times10^{43}$&0.027&$2.00\times 10^{-3}$&\\
-1.26&$7.6\times10^{44}$&$7.99\times10^{43}$&0.025&$1.87\times10^{-3}$&\\
-1.93&$1.39\times10^{44}$&$1.62\times10^{43}$&0.012&$5.76\times10^{-4}$&\\
-2.27&$0.56\times10^{44}$&$5.97\times10^{42}$&0.007&$2.24\times10^{-4}$&\\
\hline
\end{tabular}
\end{table}
The line luminosities of the major BLR lines: H$\alpha$, H$\beta$, He II, and Mg II, obtained from the realistic models of the line-emitting region, as a function of the Eddington ratio, are displayed in Fig. \ref{fig:flux_noda}. We found that all the broad lines become fainter as the accretion rate decreases, indicating the change in the AGN type. 
The type classification based solely on the presence or absence of the broad line component depends on the line to continuum contrast (or equivalent width). Since the line luminosity to the monochromatic continuum luminosity ratio is approximately constant in our models (see Fig. \ref{fig:both_intensity_noda} and \ref{fig:flux_noda}), we need to consider additional continuum component present in the observations. In Mrk 1018, \cite{Veronese2024} showed that during faint state, when the source was dimming, the V magnitude remained on the constant level, while the UV part was further fading. This is consistent with
the V band being dominated by the starlight
when the accretion disk continuum is in the faint state. Thus the broad lines have disappeared due to big drop in the line luminosities and continuum being dominated by the constant starlight component in the faint state.
These results from the simulation are in line with the fact that the AGN is transitioning from Type 1 to Type 2. This transition is also evident when comparing the BLR line luminosities between these extreme states of Mrk 1018. 
As seen in Fig. \ref{fig:flux_noda}, the quantitative drop in the line luminosities between these transitions is summarized as follows: (i) H$\alpha$ by a factor of $\sim 21$, (ii) H$\beta$ by a factor of $\sim 26$, (iii) Mg II by a factor of $\sim 24$, and (iv) He II by a factor of $\sim 10$. This dependence of BLR line strengths on the overall change of the SEDs is consistent with the observational results of the CL phenomenon (see Section \ref{sec:discussion_conclusions} for a detailed discussion), where broad lines are found to become weaker with the decrease in the accretion rate. Our result of the H$\beta$ line
flux becoming the most variable is consistent with the conclusion
of \citet{Zeltyn2024}, where the authors showed that the H$\beta$ variability is the most as compared to other lines in their sample of CL AGN.
\subsection{Fe K$\alpha$ emission models}
Next, we explore the models of the Fe K$\alpha$ emitting region to understand how the strength of Fe K emission changes with the intrinsic properties of the accretion flow, specifically the change of the mass accretion rate in the case of Mrk 1018.
Although a particular emphasis is given to the emission from the H-like and He-like ions of Fe, a general consideration of contributions of the various emission components is also discussed within the limitations of our model description. The discussion of highly ionized Fe K emission models for general SED cases in Section \ref{sec:results} clearly indicates that this emission region is much closer to the central source than the BLR. We also know,
from section \ref{sec:results}, that, a higher degree of ionization is required to produce the emission components from highly ionized He-like and H-like ions of Fe
As described in Eq. \ref{eq:U}, the ionization parameter $U$ depends on three quantities: the number of hydrogen-ionizing photons $Q(\rm H)$ [s$^{-1}$], the inner radius, and the hydrogen number density. Since $Q(\rm H)$ is constant for a given SED, there are two possible ways to increase $U$: either by decreasing the hydrogen number density or by reducing the radius of the Fe K$\alpha$ emitting region. In the first case, we assume that the Fe K emitting region has a number density of $n_{\rm H}=10^{12}$ cm$^{-3}$ consistent with the assumption in the previous section. We can now increase the ionization parameter by reducing the Fe K emitting radius, $R_{\rm FeK}$ of the emission region compared to the BLR radius $R_{\rm BLR}$, as obtained from the relation in Eq. \ref{eq:Bentz}. 

\begin{figure*}
    \centering
    \includegraphics[scale=0.5]{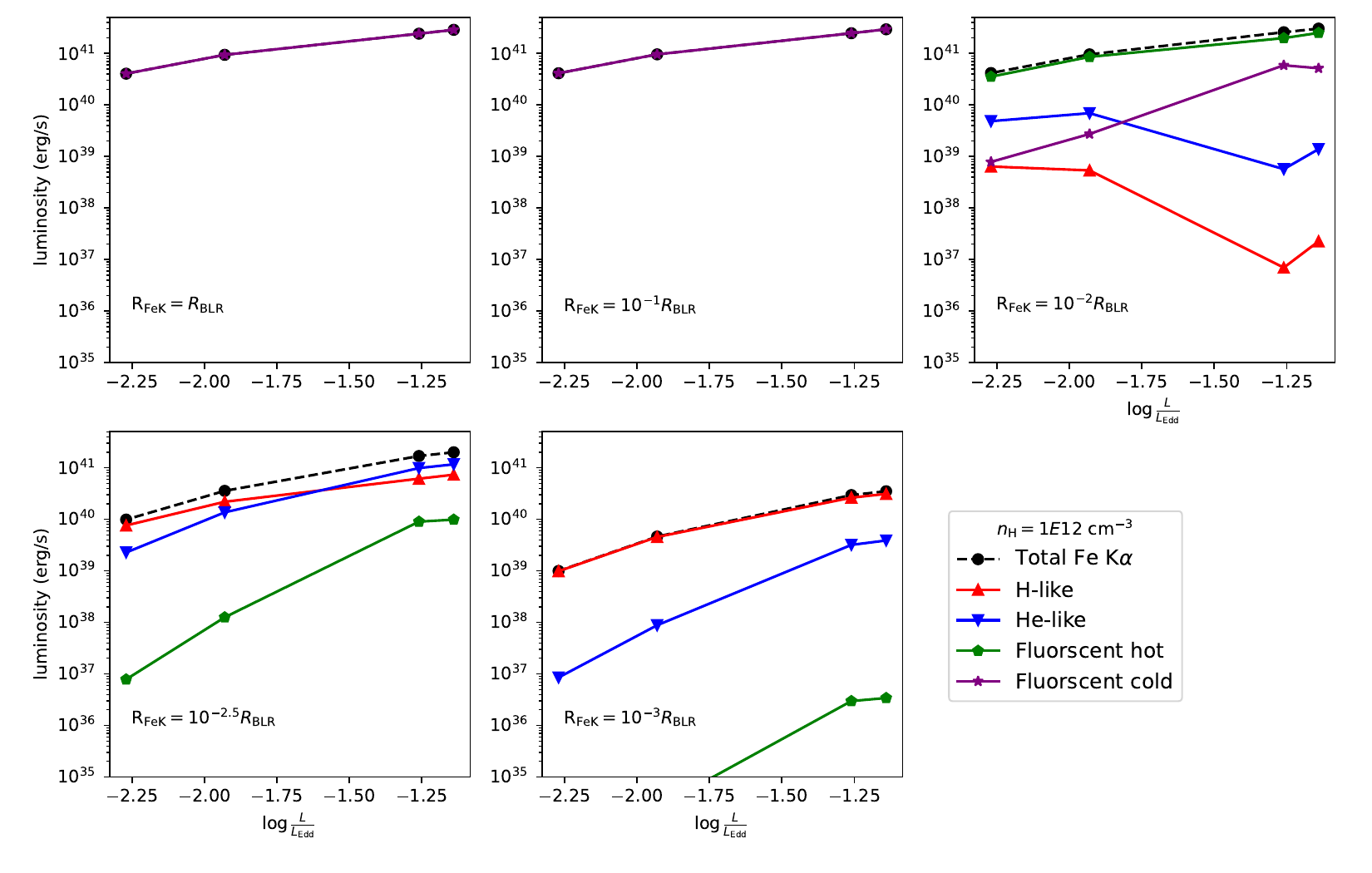}
    \caption{Plot of Fe K$\alpha$ emission luminosities for Mrk 1018 SEDs, as shown in Fig. \ref{fig:sed_noda}. The emission models are computed for a fixed gas density of $n_{\rm H} = 10^{12}$ cm$^{-3}$, with varying locations of $R_{\rm FeK}$ expressed in terms of $R_{\rm BLR}$.}
    \label{fig:Fek_noda_many_nH12}
\end{figure*}

Fig. \ref{fig:Fek_noda_many_nH12} shows the Fe K$\alpha$ emission computed with different $R_{\rm FeK}$ values, obtained by moving the emitting region closer to the central source. The first panel of the plot shows the Fe K emission from a gas cloud at the radius $R_{\rm FeK} = R_{\rm BLR}$. As expected, we do not observe any emission from higher-order Fe ions, except for the cold emission contributed by Fe ions ionized to less degree than Fe XVII. This is because the ionization is not high enough to ionize the Fe atom to the higher levels required for this contribution. Furthermore, when $R_{\rm FeK}$ decreases to orders of magnitude less than $R_{\rm BLR}$, we begin to notice contributions from higher-order He-like and H-like ions, and cold fluorescence becomes insignificant. If $R_{\rm FeK}$ further decreases to $2.5$ orders of magnitude less than $R_{\rm BLR}$, the total Fe K emission is fully dominated by the emission from He-like and H-like Fe ions. This clearly indicates that the highly ionized Fe K$\alpha$ emission is produced in the coronal region of the accretion disk in AGNs.

In the bottom panels of Fig. \ref{fig:Fek_noda_many_nH12}, where the contribution of the H-like and He-like Fe ions is dominant, we observed that the emission increases with the rise in the Eddington ratio. This means that when the AGN is identified as Type 1, the high-resolution X-ray spectra of such sources can exhibit the highly ionized Fe K$\alpha$ component. However, at the lowest Eddington accretion mode, i.e., when it is identified as Type 2, the flux of the highly ionized Fe K$\alpha$ line is lower by an order of magnitude. This suggests that the high-resolution X-ray spectra of Type 2 AGN may show a weak  Fe K$\alpha$ component, which could be detected in observations. This result is consistent 
with the conclusion of \citet{Ricci2014} who have shown, using the results of a large Suzaku study, that Seyfert 2 AGNs have on average lower Fe K$\alpha$ luminosities than Seyfert 1 for the same $10-50$ keV continuum luminosity. 
Moreover, we found that the highly ionized emission decreases as the emitting region is moved closer to the central source. This trend is evident in the lower panels of the figure, where a comparison of the plots with $R_{\rm FeK} = 10^{-2.5} R_{\rm BLR}$ and $10^{-3} R_{\rm BLR}$ demonstrates that the H-like and He-like emission decreases by an $\sim$ order of magnitude and by $\sim$ two orders of magnitude, respectively. This is expected as the H-like and He-like ion populations decrease when the Fe atom becomes fully ionized, and the medium becomes more transparent to the hard radiation field.

\begin{figure*}
    \centering
    \includegraphics[scale=0.5]{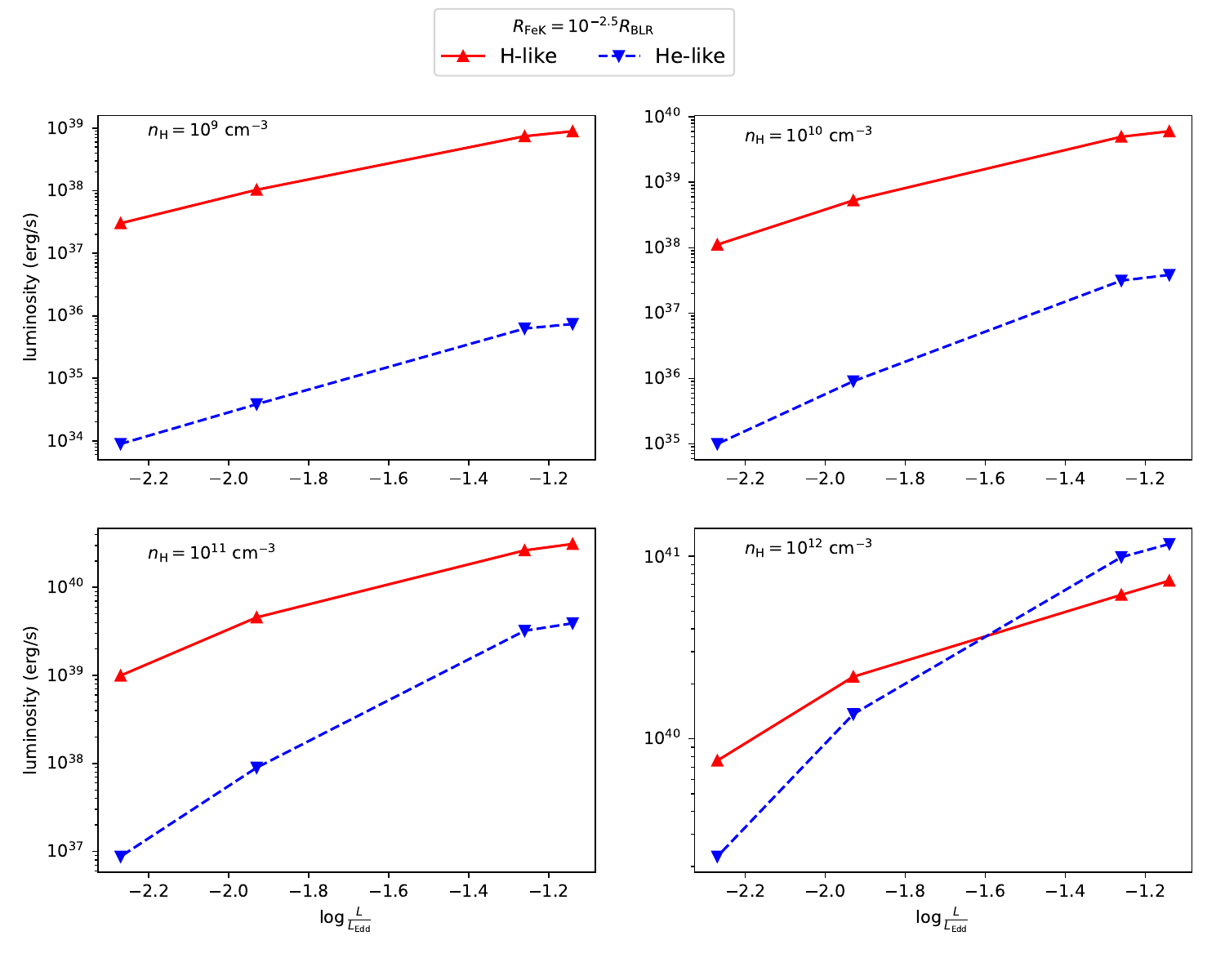}
    \caption{Luminosities of H-like and He-like Fe K emissions, plotted for various gas densities, located at the Fe-emitting radius $R_{\rm FeK}=10^{-2.5} R_{\rm BLR}$.}
    \label{fig:Fek_noda_many_nH}
\end{figure*}

Additionally, we investigated the dependence of the Fe K emission on gas density by locating the emitting region at a radius of $R_{\rm FeK} = 10^{-2.5}R_{\rm BLR}$. Fig. \ref{fig:Fek_noda_many_nH} shows the fluxes of He-like and H-like Fe emission at different stages of accretion, plotted for various gas densities ($n_{\rm H} = 10^{9}$, $10^{10}$, $10^{11}$, and $10^{12}$ cm$^{-3}$). For a given $\frac{L}{L_{\rm Edd}}$, as the gas density increases, both the He-like and H-like fluxes increase significantly. This complex dependence of the highly ionized Fe K flux on density and the location of the emitting region is also shown in Appendix \ref{sec:appendix}, in Figs. \ref{fig:Fek_noda_many_nH9}, \ref{fig:Fek_noda_many_nH10}, and \ref{fig:Fek_noda_many_nH11}. In Fig. \ref{fig:Fek_noda_many_nH9}, we found that at the low density of $10^{9}$ cm$^{-3}$, a significant Fe K flux from He-like and H-like ions can occur even at distances as far as $R_{\rm FeK} = 10^{-1} R_{\rm BLR}$.  

\section{Discussion and Conclusions}
\label{sec:discussion_conclusions}

We developed models of line-emitting regions in AGN and considered a scenario when the AGN SED and therefore ionizing radiation field changes. The changes in the ionizing radiation field are considered in two different ways:
(i) the case when only the X-ray shape of the irradiation changes, and
(ii) the case when the irradiation field across the broad energy band undergoes significant variations, in shape and the strength.

The first case of changing X-ray irradiation is achieved by varying $\alpha_{\rm ox}$ while keeping the accretion disk emission constant. In the second scenario, we consider a realistic transition of an AGN from Sy 1 type towards Sy 2 by taking the example of the well-studied CL AGN Mrk 1018. As reported by ND18, the continuum radiation of Mrk 1018 undergoes dramatic variations in the accretion flow properties, which are reflected in the changing disk and X-ray power-law emission. 
For each of these cases, we utilized the photo-ionization code {\sc cloudy}
for modeling the fluxes of lines originating in the BLR and the Fe K$\alpha$
emission region. Below, we discuss the results from each of these simulations in light of the recent observational findings regarding the association of continuum variations with changes of the line properties.

The correlation between continuum variability and BLR emission of a number of AGN
sources have been studied extensively in the literature. \citet{Denney2014} studied the changing-look (CL) AGN Mrk 590, which showed a decrease in the continuum luminosity by $\sim 2$ orders of magnitude and a simultaneous disappearance of the H$\beta$ emission line, along with the presence of a weak component of H$\alpha$ during its transition from Sy1 to Sy2. 
In the case of the CL AGN Mrk 1018, which changed its type from Sy1 to Sy1.9 between 1984 and 2006, the disappearance of the broad H$\alpha$ emission line was accompanied by a decrease in X-ray flux \citep{LaMassa2017}.
Similarly, the disappearance of the broad emission lines in the case of Mrk 590, concurrent with the $\sim 2$ orders of magnitude drop in the nuclear luminosity, was observed by \citet{Mathur2018}.
\citet{Kollatschny2020} studied the AGN IRAS 23226-3843, which showed a clear transition from Sy1 to Sy1.9 over the course of observations from 1999 to 2017, exhibiting X-ray continuum flux variations by a factor of 35, while the broad lines (H$\alpha$, H$\beta$) fluxes varied by a factor of 2. 
The AGN NGC 2992 showed a transition from 1978 (Sy2) to 2021 (Sy1), exhibiting a direct correlation of H$\alpha$ flux with the $2-10$ keV X-ray continuum \citep{Guolo2021}.

While there have been a number of reports of the CL phenomenon with respect to the H$\alpha$ and H$\beta$ lines in the past, the responsivity of the Mg II line in CL AGN is much rarer \citep{MacLeod2016}. \citet{Guo2019} reported a case of SDSS J152533.60+292012.1, which showed the turning off of the Mg II emission over a period of 286 days in tandem with only a slight variation in the optical continuum. The authors did not discuss whether this event was associated with any X-ray continuum variations. However, for the parameter space we studied here, we noticed that the Mg II flux is sensitive to changes in the strength of the continuum radiation field in both cases considered. This feature is clearly visible in Figs. \ref{fig:Lines_optical} and \ref{fig:flux_noda}. Very recently, \citet{Guo2024} conducted a systematic search for CL AGN by cross-matching spectra from the Dark Energy Spectroscopic Instrument and the Sloan Digital Sky Survey, identifying 56 CL AGN primarily based on major lines such as H$\alpha$, H$\beta$, Mg II, and other narrow emission lines. The authors also proposed 44 new possible CL AGN candidates, with significant flux variations in their broad emission lines. Eight of the sources in their sample showed simultaneous variations in more than one broad emission line. This result aligns with the trends seen in Fig. \ref{fig:flux_noda}, where we demonstrate a similar variation in different emission lines as the overall continuum changes.
All of these results are consistent with the explanation that a change in the accretion event is occurring during these phenomena in the aforementioned sources. A similar conclusion was drawn by \citet{Zeltyn2024}, where a majority of 
their CL AGN sample was found to be linked with the variations in the accretion flow. Our findings also support the view that changes in the accretion properties are likely driving the CL phenomenon in these AGNs.

From the results discussed in Section \ref{sec:results}, we can fairly argue that variations in the X-ray power-law continuum alone are unlikely to cause the strong changes in the optical line fluxes. Although we found small variations in the fluxes of all the considered BLR lines, the magnitude of these variations is always small, making them implausible for driving 
the appearance or disappearance of lines in the optical/UV spectrum. Such examples of the extreme variations in X-ray flux with the unchanged UV flux has been reported for a number of weak line quasars \citep{Ni2020ApJ,Pu2020,Zhang2023}. However, in a scenario of changing Eddington ratio, as discussed in Section \ref{sec:noda_results}, the BLR fluxes can change dramatically when the broad band shape and luminosity of the incident continuum vary. This scenario is consistent with the results of \citet{Oknyansky2019MNRAS}, where broad optical line fluxes were found to increase with the increase in the X-ray continuum and the brightening of UV/optical flux. Nevertheless, the factor by which the line fluxes change in our models is considerably higher than that reported in the observational results mentioned. The reason for this overestimation could be mainly due to the assumptions adopted in these {\sc cloudy} simulations—particularly, the identical ionization parameter for different BLR lines. We plan to address this issue in future work.

Our next aim was to understand the behavior of the highly ionized Fe K$\alpha$ line with changing broad-band SEDs in the timescale of CL phenomenon. 
Due to the limitations of the spectral resolution of X-ray instruments, it is not easy to detect the highly ionized Fe K component arising from the He-like and H-like ions of the Fe atom. In most sources where Fe K$\alpha$ is detected, it is typically dominated by cold fluorescence originating from the torus. However, there have been limited studies on the highly ionized Fe K emission originating in the coronal region of the accretion disk and its response to changes in the source continuum emission. \citet{Ballantyne2002} showed that at higher illuminating fluxes, the He-like Fe K$\alpha$ line at $6.7$ keV dominates the line complex, and this dominance depends on the ratio of the X-ray continuum flux to the accretion disk flux.
Based on a study of over 1000 X-ray sources in the XMM-COSMOS field, \citet{Iwasawa2012} found that high-ionization lines of Fe K (Fe XXV and Fe XVI) are pronounced with the increasing accretion rates.
\citet{Murphy2007} discovered a surge in the flux of highly red-shifted Fe K$\alpha$ in tandem with a dramatic increase in the $2-10$ keV continuum flux in the CL AGN NGC 2992.
Moreover, a broad component of Fe K$\alpha$ emitted in the innermost regions of the accretion disk was discovered in the same source by \citet{Marinucci2018}. Our findings of an increase in the highly ionized Fe K$\alpha$ emission with a rise in the X-ray flux of the incident continuum for a suitable model of Fe K emission are in good agreement with these observational results. We note that both of our scenarios: (i) a constant disk continuum with a change in the X-ray power-law, and (ii) a change in the broad-band continuum including the disk emission, reproduce these observed trends in coronal emission. This suggests that there could be other possibilities, beyond a change in the accretion mechanism, that could influence the Fe K$\alpha$ emission. 
A similar scenario of flaring episodes in the inner accretion disk, 
without switching off the accretion mechanism, was discussed by \citet{Murphy2007}.

Although our model description of Fe K$\alpha$ emission is primarily suitable for emission dominated by highly ionized ions of Fe, the results concerning less ionized narrow Fe K$\alpha$ emission are also interesting. As evident from Figs.~\ref{fig:Fek_noda_many_nH12}, \ref{fig:Fek_noda_many_nH9}, \ref{fig:Fek_noda_many_nH10}, and \ref{fig:Fek_noda_many_nH11}, when $R_{\rm FeK} = R_{\rm BLR}$, the total Fe K$\alpha$ luminosity is always dominated by fluorescent cold emission, regardless of the density of the emitting region. This result is consistent with the findings of \citet{Noda2023}, who discovered that in the case of CL AGN NGC 3516, the narrow Fe K$\alpha$ emitting region is co-spatial with the BLR. Moreover, the trend of narrow Fe K$\alpha$ flux variations with changes in the X-ray continuum flux is similar, showing increasing line flux with rising continuum flux at $R_{\rm FeK} = R_{\rm BLR}$. Nevertheless, a comparison of narrow Fe K$\alpha$ flux variations in our model is limited by our simplified assumptions, made to simulate Fe K$\alpha$ emission contributed primarily by H-like and He-like ions of Fe.

In both scenarios, we demonstrated that the highly ionized Fe K$\alpha$ emitting region requires an ionization parameter higher than the BLR for the emission to be noticeable. This means that the highly ionized Fe K emitting region is closer to the central source and experiences a more intense radiation field than the BLR. By comparing the two cases considered here, we found that the BLR emission is sensitive to the disk continuum, while the Fe K$\alpha$ emission is sensitive to the X-ray power-law.
The reason for this dependence is that the Fe K$\alpha$ producing region primarily receives ionizing radiation from the X-ray corona, which is located close to it, while the BLR is exposed to the softer radiation field originated from the more distant part of the disk. Our results also suggest that the corona may evolve during the CL phase of the AGN, and this evolution could be traced by detecting the changes in the highly ionized Fe K$\alpha$ emission using the high resolution X-ray observations in the future. 

Based on the {\sc cloudy} models of BLR and Fe K$\alpha$ emitting regions, we summarize our conclusions as follows:
\begin{itemize}
    \item The changes in the BLR line fluxes strongly depend on the fluctuations of the disc emission. Variations of the X-ray power-law alone do not adequately explain the appearance and disappearance of the broad-lines
    during a CL phenomenon, although a weak dependence could exist.
    \item All the major BLR lines—H$\alpha$ $\lambda 6562.80~\AA$, H$\beta$ $\lambda 4861.32~\AA$, Mg II $\lambda 2798.00~\AA$, and He II $\lambda1640.41~\AA$ showed a significant variations in flux, lowering with decreasing  Eddington ratio when an exemplary AGN changes its type from Sy1 to Sy1.9.
    \item The likely origin of the Fe K$\alpha$ line, due to emissions from H-like and He-like ions of the Fe element, is in the highly ionized coronal regions located closer to the continuum-emitting central source than in the BLR.
    \item At a given radius of the emitting region, we found a dependence of highly ionized Fe K$\alpha$ flux on the gas density, increasing with the increasing density of the medium.
    \item The highly ionized Fe K$\alpha$ flux displays flux variability similar to that of the BLR line flux variations, both in tandem with fluctuations in the irradiating continuum shape and strength.
    \item A scenario involving a change in the intrinsic properties of the accretion disk, specifically the mass accretion rate, is likely driving the CL phenomenon over timescales of months to years.
\end{itemize}
With increasing resolution of new X-ray spectrometers as micro-calorimeters on the board of X-ray Imaging and Spectroscopy Mission 
(XRISM)\footnote{https://www.xrism.jaxa.jp/en/}
and future New Advanced Telescope for High-ENergy Astrophysics (NewATHENA)\footnote{\url{https://www.esa.int/Science_Exploration/Space_Science/NewAthena_factsheet}}
mission, observations of iron line complex will put better 
constraints on its profile and variability allowing to trace 
intrinsic properties of an accretion disk. 
\begin{acknowledgments}
We thank the anonymous referee for their valuable suggestions and comments, which have greatly helped improve the content of the manuscript.
ZH and TPA acknowledges the support of the National Natural Science 
Foundation of China ( grant nos. 12222304, 12192220, and 12192221).
Some of the model computations for this article have been performed using computer 
cluster at the Nicolaus Copernicus Astronomical Center 
of the Polish Academy of Sciences (CAMK PAN). TPA greatly acknowledges CAMK PAN for this support. SM acknowledges the support from Ramanujan Fellowship
grant (RJF/2020/000113) by the Govt. of India for this research.
AR was supported by the Polish National Science Center grant No. 2021/41/B/ST9/04110.
\end{acknowledgments}

\software{astropy \citep{2013A&A...558A..33A,2018AJ....156..123A},  
          {\sc cloudy} \citep{Gunasekera2023,Ferland2017,Chatzikos2023} 
          }
\appendix
\section{Additional plots}
\label{sec:appendix}
The additional plots relevant to the presentation and discussion of the results in the main sections of the manuscript are included in this appendix.

\begin{figure}[h!]
    \centering
    \includegraphics[scale=0.55]{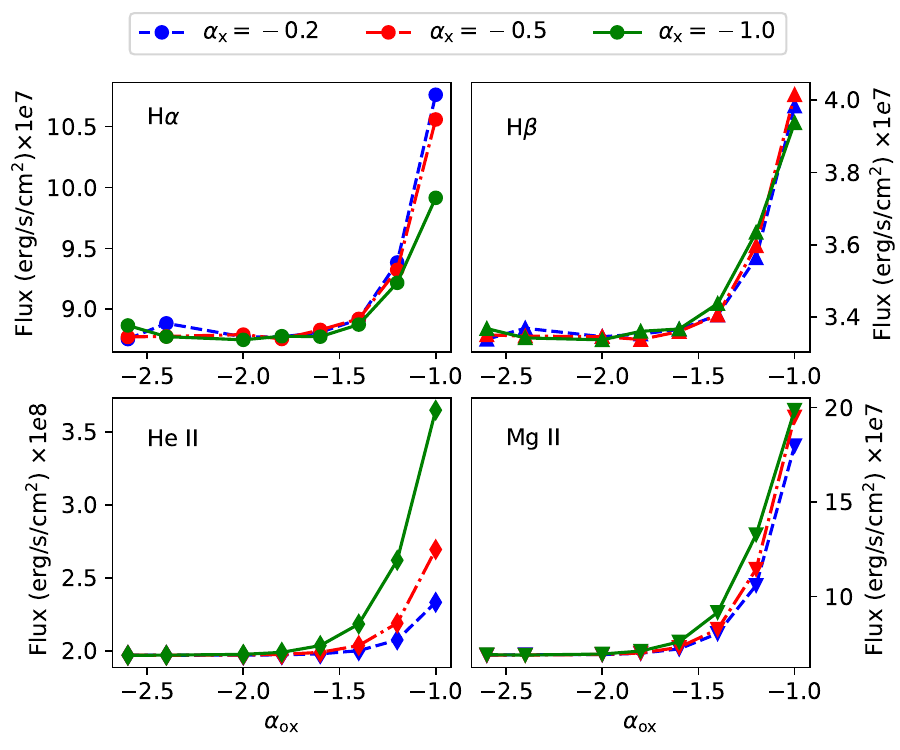}
    \caption{Fluxes for the
    considered optical/UV lines, simulated using a
    standard BLR model with $n_{\rm H}=10^{12}$ cm$^{-3}$, and $\log U = -2.0$, irradiated with SEDs corresponding to various values of $\alpha_{\rm x}=-0.2$, $-0.5$ and $-1.0$ respectively. All other model parameters are identical to those used in Fig.\ref{fig:Lines_optical}. Each panel corresponds to a specific emission line, as labeled.}
    \label{fig:optical_flux_alphaX}
\end{figure}
\begin{figure}[h!]
    \centering
    \includegraphics[scale=0.6]{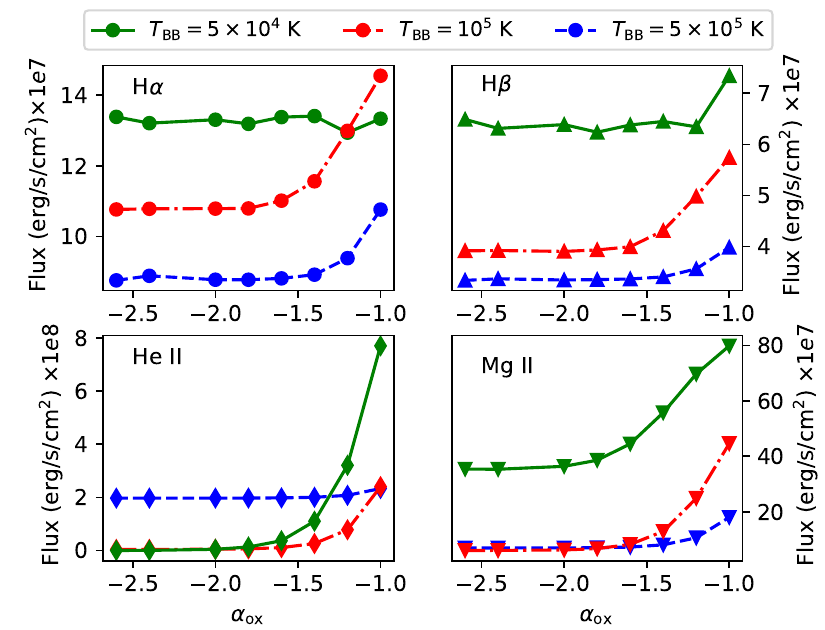}
    \caption{Fluxes for the
    considered optical/UV lines, simulated using a
    standard BLR model with $n_{\rm H}=10^{12}$ cm$^{-3}$, and $\log U = -2.0$, irradiated with SEDs corresponding to various values of disk
    temperature $T_{BB}=5\times 10^{4}$, $10^{5}$ and $5\times10^{5}$ K respectively. All other model parameters are identical to those used in Fig.\ref{fig:Lines_optical}. Each panel corresponds to a specific emission line, as labeled.}
    \label{fig:optical_flux_var_T}
\end{figure}
\begin{figure*}[h!]
    \centering
    \includegraphics[scale=0.6]{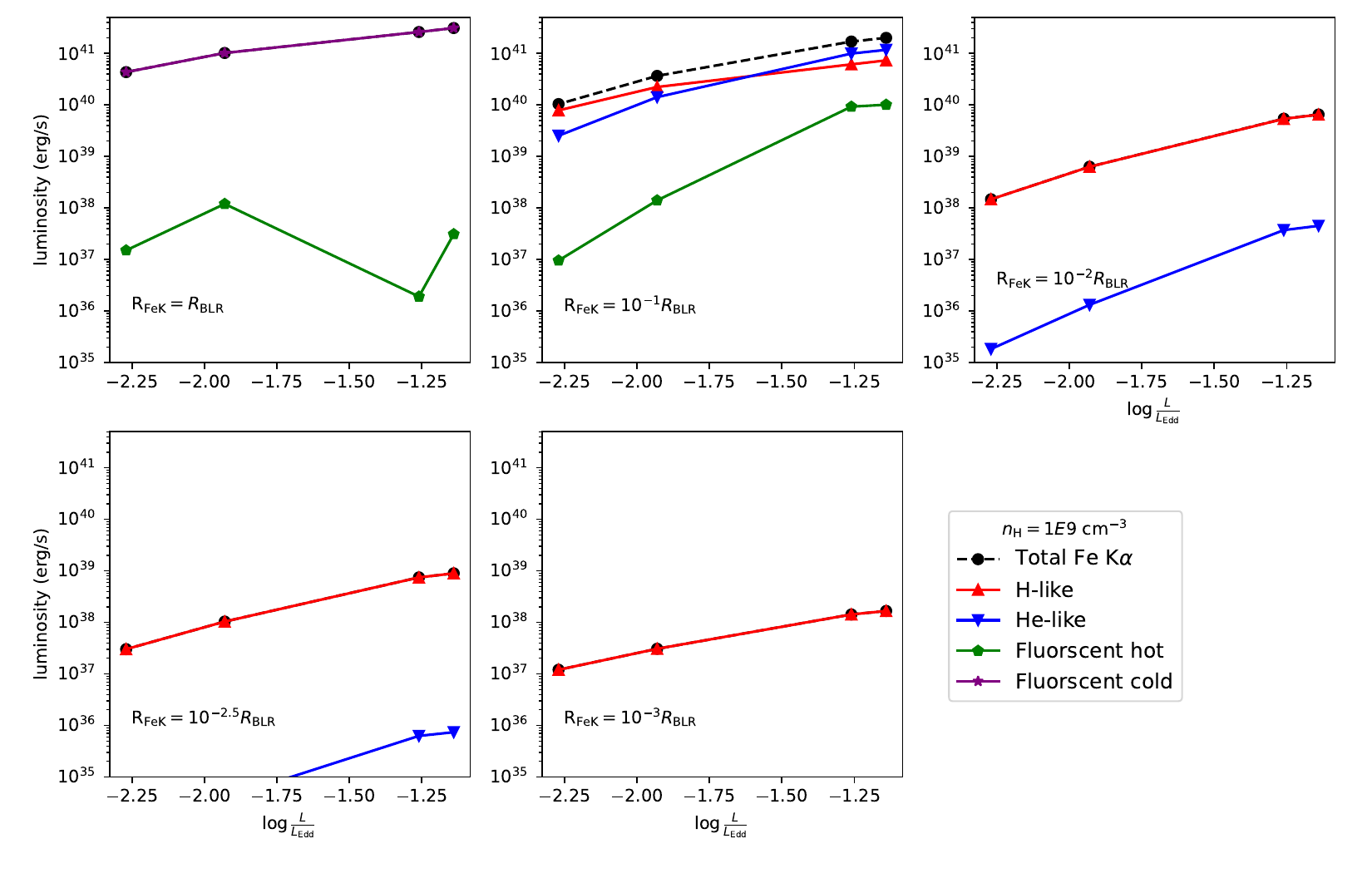}
    \caption{Plot of Fe K$\alpha$ emission luminosities for Mrk 1018 SEDs, as shown in Fig. \ref{fig:sed_noda}. The emission models are computed for a fixed gas density of $n_{\rm H} = 10^{9}$ cm$^{-3}$, with varying locations of $R_{\rm FeK}$ expressed in terms of $R_{\rm BLR}$.}
    \label{fig:Fek_noda_many_nH9}
\end{figure*}

\begin{figure*}[h!]
    \centering
    \includegraphics[scale=0.6]{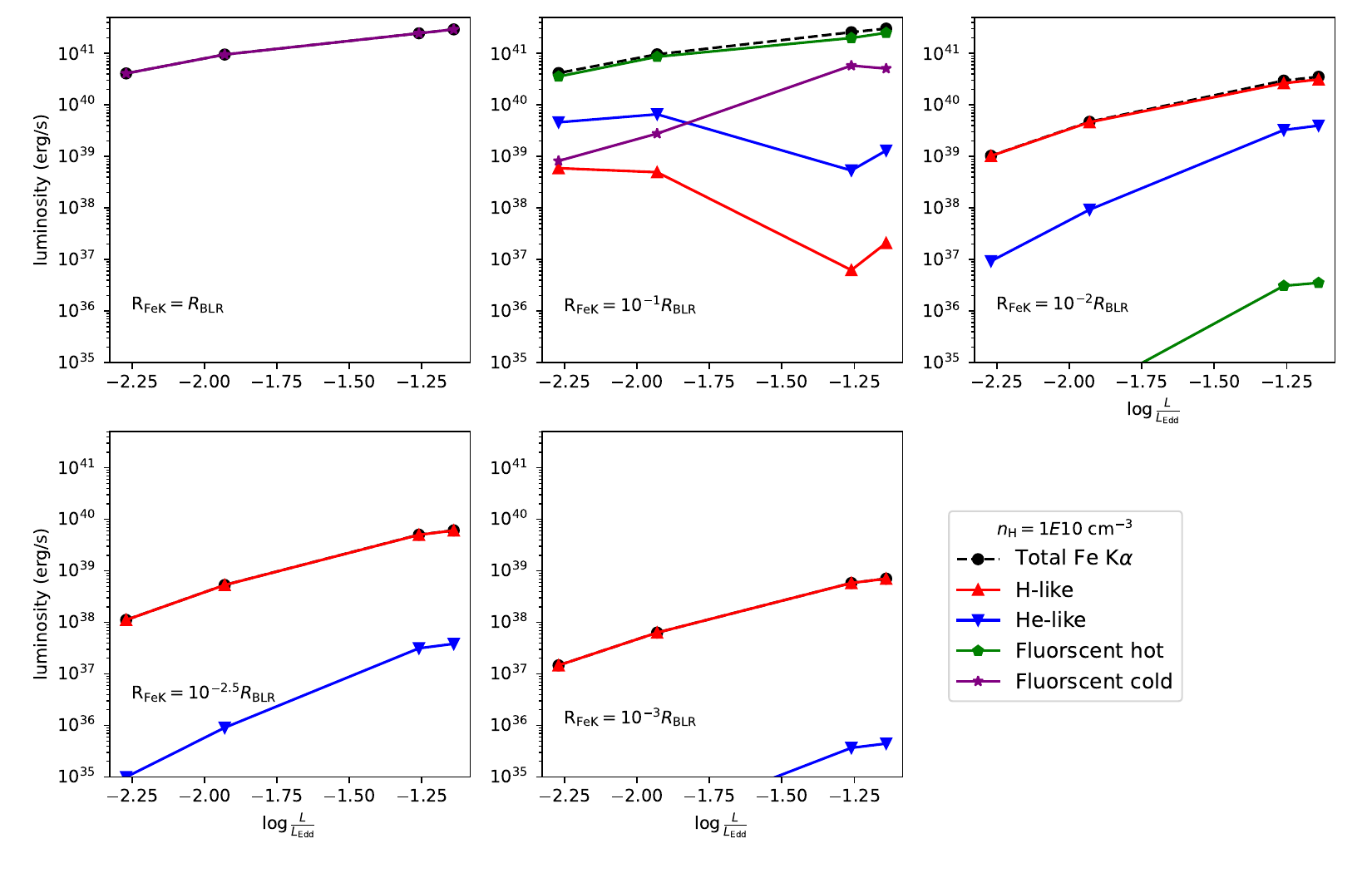}
    \caption{Similar to Fig. \ref{fig:Fek_noda_many_nH9} plotted for the emitting models with $n_{\rm H} = 10^{10}$ cm$^{-3}$}
    \label{fig:Fek_noda_many_nH10}
\end{figure*}

\begin{figure*}[h!]
    \centering
    \includegraphics[scale=0.6]{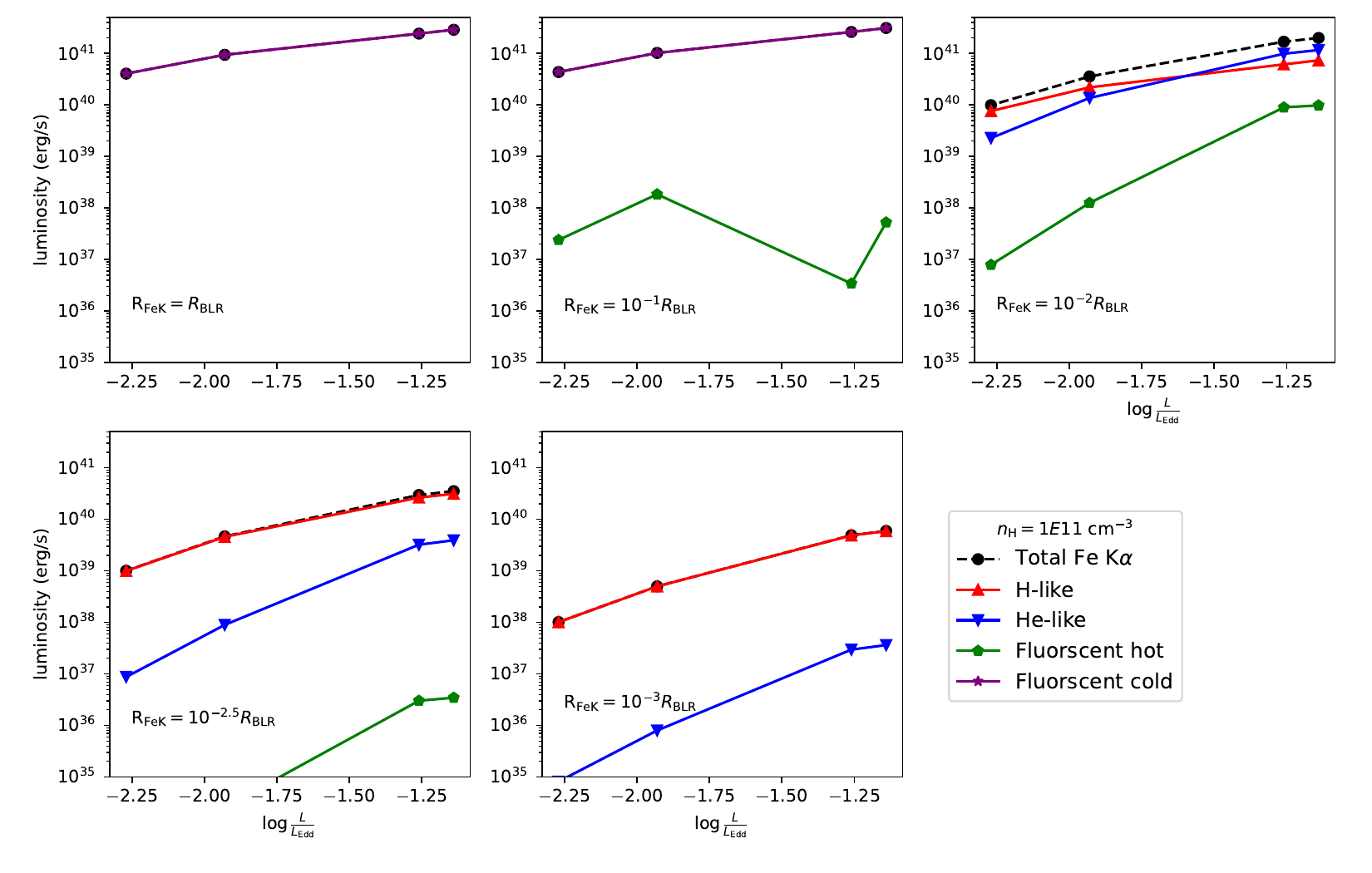}
    \caption{Similar to Fig. \ref{fig:Fek_noda_many_nH9} plotted for the emitting models with $n_{\rm H} = 10^{11}$ cm$^{-3}$}
    \label{fig:Fek_noda_many_nH11}
\end{figure*}

\bibliography{references}{}
\bibliographystyle{aasjournal}
\end{document}